%% file: main.tex
\pdfoutput=1
\documentclass[aps,prl,twocolumn,superscriptaddress,10pt]{revtex4-2}
\usepackage{amsmath,amssymb,graphicx,bm}
\usepackage{hyperref}
\usepackage{physics}
\usepackage{etoolbox}
\usepackage[version=4]{mhchem}
\usepackage{siunitx}

\makeatletter
\let\frontmatter@footnote@produce\frontmatter@footnote@produce@footnote
\patchcmd{\doauthor}
  {\ignorespaces#1\unskip\@listcomma}
  {\ignorespaces#1\unskip}
  {}{}
\patchcmd{\doauthor}
  {\space \@listand}
  {\@listcomma\space \@listand}
  {}{}
\makeatother

\newcommand{\yb}{\ce{^{171}Yb^+}}
\DeclareSIUnit\gauss{G}
\DeclareSIUnit{\rad}{rad}
\DeclareSIUnit{\watt}{W}
\AtBeginDocument{\RenewCommandCopy\qty\SI}

\begin{document}

\title{All-electronic geometric phase gate on Doppler-cooled trapped ions using an axial mode}

\author{P. Rahimi}
\affiliation{Sussex Centre for Quantum Technologies, University of Sussex, Brighton, BN1 9QH, UK}

\author{C. Knapp}
\affiliation{Sussex Centre for Quantum Technologies, University of Sussex, Brighton, BN1 9QH, UK}

\author{P. Zantis}
\affiliation{Sussex Centre for Quantum Technologies, University of Sussex, Brighton, BN1 9QH, UK}

\author{T. Maddock}
\affiliation{Sussex Centre for Quantum Technologies, University of Sussex, Brighton, BN1 9QH, UK}

\author{S. J. Hile}
\affiliation{Sussex Centre for Quantum Technologies, University of Sussex, Brighton, BN1 9QH, UK}

\author{J. Randall}
\affiliation{Universal Quantum Ltd, Gemini House, Mill Green Business Estate, Haywards Heath, RH16 1XQ, UK}

\author{M. Siegele-Brown}
\affiliation{Sussex Centre for Quantum Technologies, University of Sussex, Brighton, BN1 9QH, UK}

\author{S. Weidt}
\affiliation{Sussex Centre for Quantum Technologies, University of Sussex, Brighton, BN1 9QH, UK}
\affiliation{Universal Quantum Ltd, Gemini House, Mill Green Business Estate, Haywards Heath, RH16 1XQ, UK}

\author{W. K. Hensinger}
\email{w.k.hensinger@sussex.ac.uk}
\affiliation{Sussex Centre for Quantum Technologies, University of Sussex, Brighton, BN1 9QH, UK}
\affiliation{Universal Quantum Ltd, Gemini House, Mill Green Business Estate, Haywards Heath, RH16 1XQ, UK}

\date{\today}

\begin{abstract}
We report the experimental demonstration of a high-fidelity oscillating magnetic gradient entangling gate on the axial stretch mode of a two-ion crystal in a micro-fabricated surface Paul trap, therefore allowing for mixed-species sympathetic cooling. With only Doppler-cooled \yb{} ions at $\bar{n}=50(5)$, we achieve Bell-state fidelities far above the relevant fault-tolerant threshold, simplifying transport protocols within a shuttling-based-architecture. The control fields consist only of an oscillating current applied to submerged integrated wires emitting a B-field gradient near-resonant with the axial mode, and resonant microwave pulses at the qubit frequency constituting very low gate calibration overheads. We also demonstrate an adiabatic elimination of spin-motion entanglement (AESE) version of the gate, showcasing significant robustness.
\end{abstract}

\maketitle

\section{Introduction}
\label{sec:intro}
Trapped atomic ions are among the most mature platforms for quantum information processing, combining naturally identical qubits, long coherence times, and high-fidelity state preparation and measurement~\cite{Bruzewicz2019,An2022,Sotirova2024,Wang2021,Pi2026}. Universal gate sets have also been successfully demonstrated, with two-qubit gate errors at or below the $10^{-3}$ level in earlier laser-driven implementations~\cite{Ballance2016, Gaebler2016}, and have since matured into larger-scale systems~\cite{Moses2023,Ransford2026}. `Laser-free' methods have also emerged as promising candidates for entanglement generation in a scalable quantum computing (QC) architecture, boasting high entanglement fidelities~\cite{Srinivas2021,Loschnauer2025} even without the need for ground-state cooling~\cite{Srinivasphd,Hughes2025}.
\par
Whilst near-ground state cooling is not strictly necessary to attain high-fidelity two- and multi-qubit entanglement~\cite{Molmer1999,Sorensen2000}, the $\sigma_x \otimes \sigma_x$ implementations common in earlier work on trapped-ion systems~\cite{Harty2016,Weidt2016,Hahn2019,Zarantonello2019} benefitted massively from low mode occupation, as many non-trivial error channels scale poorly with temperature~\cite{Sutherland2022}. As recent demonstrations have also shown the sideband cooling sequence to be a limiting factor for clock speeds in quantum charge-coupled device (QCCD) architectures~\cite{Pino2021,Moses2023}, eliminating the need for ground-state cooling is of critical importance for a prospective large-scale system~\cite{Kielpinski2002,Akhtar2023}~\textemdash~not only with regard to the efficient implementation of quantum algorithms through reducing the number of control fields and experimental complexity, but also as a means of reducing the energy cost of operating the device~\cite{carrascocodina2026} and significantly enhancing its clock speed.
\par
To this end, efforts have been made to increase robustness to residual spin-motion entanglement errors via coherent control methods~\cite{Haddadfarshi2016,Webb2018,Shapira2018,Valahu2022}, though these often come at the cost of both longer operations and greater experimental complexity i.e.~engineering of the control sequence to completely disentangle the spin and motion at the end of the process which, critically, assumes the system parameters will not fluctuate. Recent work has explored adiabatically eliminating spin-motion entanglement (AESE), in which the spin-dependent force is smoothly ramped with respect to the ions' motion~\cite{Sutherland_2024,Hughes2025}, relaxing experimental control requirements and providing a pathway toward hot entangling gates.
\par
Additionally, when viewed from a broader perspective, to realise a useful-scale QCCD one must also evaluate a potential entanglement generation method in the context of the architecture's demands~\textemdash~not only will the entangling gates be required to exhibit robustness to mode frequency fluctuations, but also demonstrate compatibility with mid-circuit cooling, robustness to shuttling operations and qubit decoherence, and present no irreducible error floors or untenable power requirements, whilst being as simple as possible to implement in terms of control fields. The seminal work of Ospelkaus et.~al.~\cite{ospelkaus_trapped-ion_2008} proposed a $\sigma_z \otimes \sigma_z$ gate which makes use of oscillating gradients near-resonant with the ions' secular frequency. This method is also compatible with dynamical decoupling sequences that can counteract qubit frequency noise, because dynamical decoupling pulses about $x$ do not disrupt the $\sigma_z \otimes \sigma_z$ interaction~\cite{Sutherland_2019}. This allows for dynamical decoupling and gate design to be separable, relaxing control requirements. Therefore, a viable avenue for a scalable entangling gate in a trapped-ion QC would be one that is performed in the $z$-basis, compatible with AESE-type sequences, and not in contrast with any of the broader requirements listed above.
\par
To implement an entangling gate on a chip without the use of lasers, a method of coupling the spin and motion must be engineered. This is commonly achieved through current-carrying wires (CCWs) on the surface of the chip~\cite{Srinivas2021,Loschnauer2025}. However, whilst fabricating the CCWs in this way reduces the manufacturing overhead, the oscillating electric field generated in close proximity to the ions subjects them to an electric force that can excite the ions' motion, leading to an array of unwanted effects which must be carefully calibrated out or mitigated~\cite{Srinivasphd}, thus rendering $\sigma_z \otimes \sigma_z$ gates challenging to implement.
\par
In this work, these restrictions are lifted through utilising a unique chip architecture in which the CCWs are buried beneath the surface of the chip~\cite{Siegele2022}, shielded from the trapping location by a grounding layer, which protects the ions from unwanted electric field coupling. We experimentally demonstrate a $\sigma_z \otimes \sigma_z$-type entangling operation, with inherent robustness to magnetic field fluctuations, and a simple experimental control sequence consisting only of resonant microwave mapping pulses and an oscillating magnetic field gradient. The gate is performed without the need for sub-Doppler cooling, resulting in an attractive scheme for a large-scale quantum computing architecture. As large-scale quantum computers will likely require mid-circuit cooling to maintain gate fidelities~\cite{Ransford2026}, use of an axial mode in this work makes this gate mechanism compatible with mixed-species sympathetic cooling, as well as other promising mid-circuit cooling methods innovations~\cite{Fallek2024, smith2026}. Axial mode coupling is also advantageous in the context of mixed-species (such as Ba-Yb) entangling gates, for the same reason that mode coupling asymmetry is far lower, and hence gate duration much shorter, as compared to coupling via a radial AC gradient. We prepare a maximally entangled Bell state with a fidelity of $100.0^{+0.0}_{-0.4}\%$, showcasing the efficacy of near-motional oscillating gradient gates in overcoming both thermal and hardware-induced sources of error. We also implement an AESE-type gate similar to that proposed in~\cite{Hughes2025}, demonstrating the trade-off between speed and motional robustness.

\section{Theory}
We demonstrate a gate mechanism first proposed by Ospelkaus et.~al.~in~\cite{ospelkaus_trapped-ion_2008}, and built upon by Sutherland et.~al.~\cite{Sutherland2022,Sutherland_2024} and Hughes et.~al.~\cite{Hughes2025}, the fundamentals of which are summarised here. For two ions confined along the axial ($z$) direction of an RF Paul trap, we consider the general, idealised two-qubit gate Hamiltonian in the laboratory frame, given by
\begin{equation}
\label{eq:lab_H}
\hat{H}_m(t)=\frac{\hbar\omega_0}{2}\hat{S}_z +
\hbar\omega_m\hat{a}^{\dagger}\hat{a} +
\hbar\Omega_g f(t)\hat{S}_{z,m}\!\left(\hat{a}^{\dagger}+\hat{a}\right),
\end{equation}
for the gating motional mode $m$. The first term represents the internal splitting of the qubit states where $\hat{S}_z=\hat{\sigma}_{z,1}+\hat{\sigma}_{z,2}$ is the total qubit operator, whilst the second denotes the energy of the selected collective motional mode of frequency $\omega_m$. The third term defines the oscillating-gradient-mediated spin-motion coupling, where $f(t)=g(t)\cos[\omega_m t-\varphi(t)]$, with $g(t)$ an arbitrary pulse envelope and $\varphi(t)=\int_0^t\delta(t')dt'$ the accumulated phase for an instantaneous drive frequency $\omega_m-\delta(t)$, where $\delta(t)$ is a time-dependent gate detuning. For a constant detuning, this reduces to $\varphi(t)=\delta t$. $\hat{S}_{z,m}=\hat{\sigma}_{z,1}\pm\hat{\sigma}_{z,2}$ is the collective mode-weighted spin operator for the centre-of-mass (COM, $+$) or stretch ($-$) axial modes. $\Omega_g$, which quantifies the strength of the longitudinal spin-dependent force on the axial motion, is given by
\begin{equation}
\label{eq:gate_rabi}
\Omega_g=\frac{1}{4}\sqrt{\frac{\hbar}{M\omega_m}}\frac{\partial B}{\partial z}\frac{\partial \omega_0}{\partial B},
\end{equation}
for two ions of equal mass $M$. The speed of the interaction therefore depends on the mass of the ion(s), the frequency of the chosen mode, the zero-to-peak amplitude of the oscillating axial gradient, $\partial B/\partial z$, and the magnetic sensitivity of the chosen qubit state(s), $\partial\omega_0/\partial B$.
\par
This third Hamiltonian term fundamentally describes a spin-dependent excitation of the shared motion, which couples both spins to the shared oscillator. For the two non-zero eigenvalues of the eigenstates of $\hat{S}_{z,m}$, the spin-states are forced in opposite phase-space trajectories, eliciting a superposition of different motional displacements. By transforming into a rotating frame with respect to the qubit splitting and motion and making the rotating wave approximation, Equation~\ref{eq:lab_H} becomes
\begin{equation}
\hat{H}_I(t)=\hbar\Omega_g' g(t)\hat{S}_{z,m}
\left(\hat{a}^{\dagger}e^{i\varphi(t)}+\hat{a}e^{-i\varphi(t)}\right),
\end{equation}
where $\Omega_g'=\Omega_g/2$ is the co-rotating half of the RF gradient drive. The exact propagator can be found via the Magnus expansion~\cite{magnus1954} and is given by
\begin{equation}
\label{eq:propagator}
\hat{U}(t)=\hat{D}\left[\alpha(t)\hat{S}_{z,m}\right]
\exp\left[\pm i\Phi(t)\hat{S}^2_{z,m}\right].
\end{equation}
The $\hat{D}\left[...\right]$ factor describes the aforementioned spin-dependent displacement corresponding to the phase-space trajectories of the spin states' motion, while the second term denotes the accumulated geometric phase. For an arbitrary pulse envelope, the displacement is
\begin{equation}
\label{eq:alpha_general}
\alpha(t)=-i\Omega_g'\int_0^t g(t')e^{i\varphi(t')}dt'.
\end{equation}

For a constant drive $g(t)=1$ and detuning $\delta(t)=\delta$, Equation~\ref{eq:alpha_general} reduces to $\alpha(t)=\Omega_g'(1-e^{i\delta t})/\delta$, such that $t_g=2\pi N/|\delta|$ gives $N$ closed phase-space circles.
Trajectory closure can be achieved by more general $g(t)$ and $\varphi(t)$ pairs by enforcing $\alpha(t_g)=0$ after the gate duration $t_g$, leaving the idealised gate propagator $\hat{U}(t_g)=\exp[\pm i\Phi(t_g)\hat{S}^2_{z,m}]$.
The geometric phase $\Phi(t)$ is in general given by the enclosed loop area
\begin{equation}
\label{eq:phi_general}
\Phi(t)=\operatorname{Im}\int_0^t\alpha^*(t')d\alpha(t').
\end{equation}

Expanding the collective spin operator yields $\hat{S}^2_{z,m}=2\hat{I}\pm2\hat{\sigma}_{z,1}\hat{\sigma}_{z,2}$. The first term contributes only a global phase, leaving $\hat{U}(t_g)=\exp[\pm 2i\Phi(t_g)\hat{\sigma}_{z,1}\hat{\sigma}_{z,2}]$, the effective $\sigma_z\otimes\sigma_z$ interaction, and generating a maximally entangling operation when $\left| \Phi(t_g)\right|=\pi/8$.
\par
As Equation~\ref{eq:propagator} describes the entanglement of spin and motion during a loop, any alteration of the phase-space trajectory via fluctuations of the motional mode frequency will affect the spins, leading to infidelity. References~\cite{Sutherland_2024} and~\cite{Hughes2025} show that adiabatic drive modulation (`AESE') and (to a lesser extent) Walsh modulation can allow for the engineering of robustness to residual spin-motion entanglement at the end of the gate operation.

\begin{figure*}[t]
    \centering
    \includegraphics[width=1\textwidth]{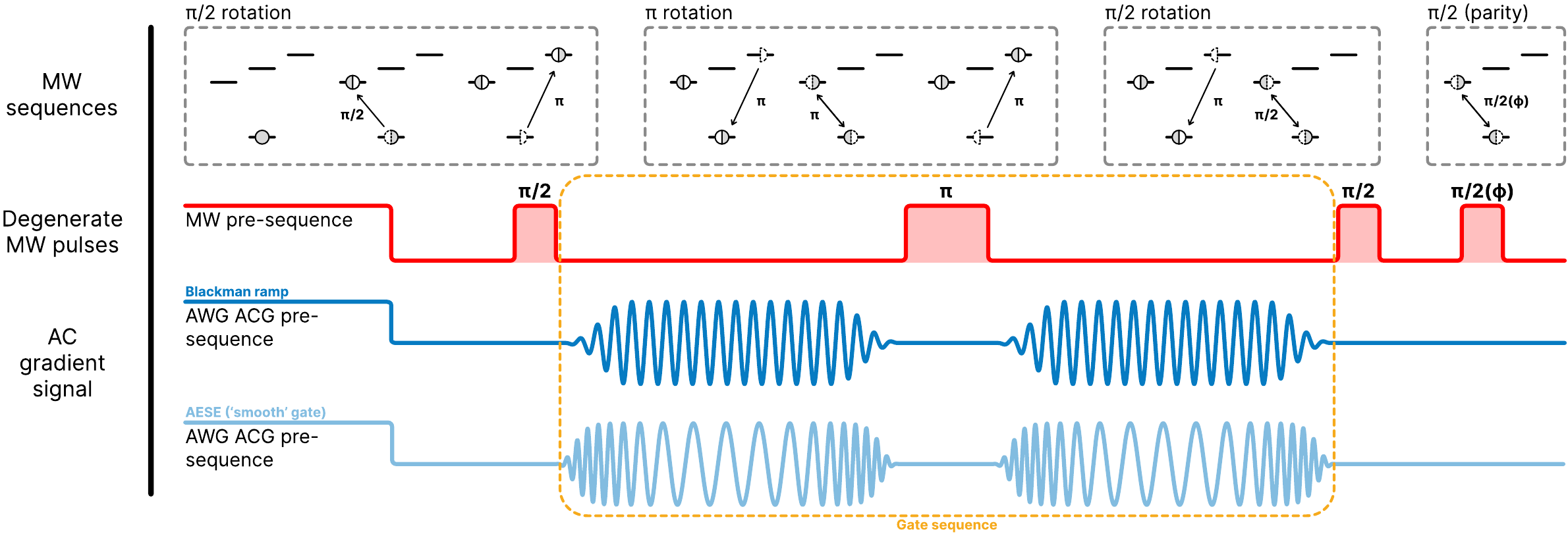}
    \caption[]{Experimental sequence for implementation of the gate. Microwave (MW) pulses (red, middle) are applied when the CCW current is zero, driving population in both ions in a Hahn echo-type sequence. The oscillating gradient (blue, bottom) is applied during the MW inter-pulse delays, each accruing half of the required geometric phase, and are ramped in amplitude with a \SI{10}{\micro\s} Blackman ramp. Due to the initial superposition state comprised of $\ket{F=1, m_F=+1}$ and $\ket{F=1,m_F=-1}$, a series of MW pulses are required to implement the desired $\pi$ and $\frac{\pi}{2}$ rotations, which are depicted in grey on the top row, illustrating how population is transferred between the states in the ground state manifold of the two \yb{} ions. The MW pre-sequence consists of Doppler cooling and state preparation, whilst the subsequent AC gradient pre-sequence contains four sequential gradient pulses, which were found to improve the amplitude stability of the signal by thermalising the amplifier more effectively over its duty cycle. The gradient waveforms are schematic and not to scale.}
    \label{fig:Pulsesequence}
\end{figure*}
\section{Experimental Setup}
\label{sec:setup}
The experiment is performed with a pair of \yb{} ions trapped at a height of $\SI{125}{\micro\meter}$ above a microfabricated surface Paul trap, cooled to approximately \SI{54}{\kelvin} via helium gas~\cite{Foni2022}. The device is indium bonded to a gold-plated copper block, which makes thermal contact to the in-vacuum helium heat exchanger via mechanical compression. The trap itself features a set of buried current-carrying wires (CCWs) fabricated approximately \SI{11}{\micro\meter} beneath the surface of the chip as detailed in~\cite{Siegele2022}. To prevent the strong excitation of motional modes via electric fields, the wires are electrically shielded from the ion trap by a \SI{1}{\micro\meter} thick Au ground plane embedded within SiO$_2$. This shielding is critical, as unshielded electric fields oscillating near the secular frequencies can cause the ions to sample anharmonic regions of the potential, or even leave the trap~\cite{srinivas_trapped-ion_2018}. The CCW signal is synthesised via arbitrary waveform generator (AWG), amplified, and impedance matched to the CCWs.
\par
The ions are confined such that the axial stretch mode frequency is $\omega_s = 2\pi \times\SI{499.63(1)}{\kilo\hertz}$. A background quantising magnetic field is set to \SI{16.8}{\gauss}, providing significant Zeeman splitting to reduce the impact of hardware-induced AC Zeeman shifts that arise during the gate (described below). The gate utilises the maximally magnetic field sensitive superposition in the $^2S_{1/2}$ ground state of \yb{}, specifically the $\ket{F=1,m_F=+1}$ and $\ket{F=1,m_F=-1}$ states, providing a differential magnetic field sensitivity of \SI{2.8}{\mega\hertz\per\gauss} which maximises the entanglement rate for a given gradient amplitude.
\par
Prior to the entangling operation, the only required calibrations are a measurement of the secular frequency, and a single magnetic-sensitive transition to measure the magnetic field at the ion. It must also be ensured that the coherence of the qubit states are sufficiently maintained over the duration of the entangling sequence, which can be verified via a spin-echo sequence with identical timings to the entanglement operation (see Supplemental Material~\cite{supp}). All sequences are line-triggered to a \SI{50}{\hertz} mains pick-off signal to reduce magnetic field noise.
\section{Gate Protocol and Error Mitigation}
\label{sec:gate}
\subsection{State preparation}
Both \yb{} ions are first Doppler cooled for \SI{15}{\milli\s}, resulting in a gate mode (axial, stretch) temperature of $\bar{n}=50(5)$~\cite{supp}. The high temperature relative to the Doppler limit of $\bar{n}_D=19.7$ is due to a broadened line-width~\textemdash~a consequence of operating at a large magnetic field. They are then optically pumped into the $\ket{^2S_{1/2},F=0}$ state using a \SI{369.5}{\nano\meter} laser driving the $\ket{^2S_{1/2},F=1} \longrightarrow \ket{^2P_{1/2},F=1}$ resonance, assisted by an electro-optic modulator (EOM) operating at \SI{2.1}{\giga\hertz}, applied for \SI{50}{\micro\s}. Following initialisation, a sequence of global microwave pulses prepares both ions in a superposition of the $\ket{F=1, m_F=+1}$ and $\ket{F=1,m_F=-1}$ states, as shown in Figure~\ref{fig:Pulsesequence}. The total duration of the qubit preparation pulses is \SI{12}{\micro\s}, and is only limited by the available microwave drive power.
\subsection{Blackman and AESE gate implementation}
\label{subsec:sequence}
The pulse sequences for the experiments performed in this work are shown in Figure~\ref{fig:Pulsesequence}. For the Blackman sequence, \SI{2.2}{\watt} (peak) is applied to the CCWs at a frequency of $\omega_m-\delta=\SI{506.9}{\kilo\hertz}$, corresponding to a detuning from the chosen axial stretch mode of $\delta=\SI{-7.25}{\kilo\hertz}$. This results in a peak axial oscillating gradient of $\partial_zB\approx\SI{35}{\tesla\per\meter}$~\cite{supp} at each ion during the gradient pulses. This work makes use of two-loop sequences in which the desired geometric phase is accumulated over the course of two gradient pulses, separated by a refocusing $\pi$-pulse sequence. This is to counteract both the induced AC Zeeman shift, and any additional slow background magnetic field noise.
\par
The resulting maximum gate Rabi frequency for the implemented gate sequence as prescribed by Equation~\ref{eq:gate_rabi} is $\Omega_{max} = \SI{2.67}{\kilo\hertz}$, with corresponding Bell state generation times of $t_{g,B}=\SI{300}{\micro\s}$ and $t_{g,AESE}=\SI{570}{\micro\s}$ for the Blackman and AESE sequences respectively. The time to create a Bell state depends on which robustness protocol is implemented, and excludes the microwave mapping pulses and short sections of padding surrounding the gradient pulses corresponding to the ring-down time of the resonance matching circuit.
\par
For the Blackman sequence, the two gradient pulses are $\SI{150}{\micro\s}$ in length, with ramping times for the gradient amplitude and detuning of $\tau_{\Omega}=\SI{10}{\micro\s}$. The amplitude is ramped from 0 to $\Omega_{max}$ and symmetrically back to 0. In the AESE sequence, the recipe outlined in~\cite{Hughes2025} is slightly modified such that the ramp of the detuning $\tau_g < t_g/2$, producing a more flat-top profile. The timing of the ramp is determined via numerical optimiser, which works to minimise the displacement $|\alpha(t_g)|$ that Equation~\ref{eq:alpha_general} leaves at the end of a pulse, over a range of offsets of the secular frequency, for a given range of amplitude and detuning ramp times. The optimiser was run with Gaussian profiles, and found that sufficient robustness to static secular frequency fluctuations can be achieved, with the benefit of maintaining relatively faster entanglement generation compared to profiles which ramp over the entire duration of the gate. For the AESE gate, the two gradient pulses are $\approx \SI{285}{\micro\s}$ in length, with ramping times for the gradient amplitude and detuning of $\tau_{\Omega}=\SI{12}{\micro\s}$ and $\tau_{\delta}=\SI{11}{\micro\s}$ respectively (10\% to 90\%). The amplitude is ramped from 0 to $\Omega_{max}=\SI{1.9}{\kilo\hertz}$, and the detuning is ramped from $\delta_{max}=\SI{-14.4}{\kilo\hertz}$ to $\delta_{min}=\SI{-6.2}{\kilo\hertz}$. Both the amplitude and gate detuning are then held for $t_g-2\tau_{\Omega}$ and $t_g-2\tau_{\delta}$ respectively, before being ramped back to $\Omega=0$ and $\delta=\delta_{max}$. The AESE sequence was also performed on the axial stretch mode.

\subsection{Projection and readout}
\label{subsec:projection}
Following the second gradient pulse, a $\frac{\pi}{2}$-rotation (the remapping pulse) is used to transfer population back to the $\{\ket{\downarrow}=\ket{F=0}, \ket{\uparrow}=\ket{F=1}\}$ readout basis in \yb{}. The phase of the remapping $\frac{\pi}{2}$-pulse is determined by finding the systematic phase accumulated during the gate pulses, and is chosen such that the amplitude of the parity oscillation is maximised. Finally, a parity $\frac{\pi}{2}$-pulse of varying phase selects the projection axis prior to fluorescence-based readout.

\subsection{Hardware induced AC Zeeman shifts}
\label{subsec:light_shifts}
We observe an amplitude dependent AC Zeeman shift of the qubit states during the gradient pulses~\cite{supp}, which can be explained by an out-of-plane magnetic field component that increases with gradient size. A likely explanation for this is the presence of eddy currents induced within the copper heat sink located beneath the chip. These eddy currents displace the quadrupole null of the oscillating gradient from the intended \SI{120}{\micro\meter} to approximately \SI{150}{\micro\meter}. Consequently, the ions are subjected to a strong out-of-chip oscillating magnetic field during the gate. At $\approx\SI{35}{\tesla\per\meter}$, and with a quantising field of \SI{16.8}{\gauss} (corresponding to a Zeeman splitting of \SI{23.5}{\mega\hertz}), this residual oscillating field off-resonantly induces an AC Zeeman shift of approximately \SI{5460}{\kilo\hertz} ($\approx\SI{2730}{\kilo\hertz}$ per $m_F = \pm 1$ state). This results in a large $\sigma_z$ phase accumulation of approximately \SI{4900}{\rad} per gradient pulse on each ion. The out-of-chip field is nearly the same at the two ions, so that most of this phase is common to them; slight mismatches in the ions' positions relative to the null and quantisation axis make it differ between the two ions by about 2\%~\cite{supp}. The Hahn echo sequence illustrated earlier is able to reverse unwanted phase accumulation on both qubits, however, cannot compensate small amplitude fluctuations in the gradient amplitude between the two pulses.
\begin{figure}[t]
  \centering
  \includegraphics[width=\linewidth]{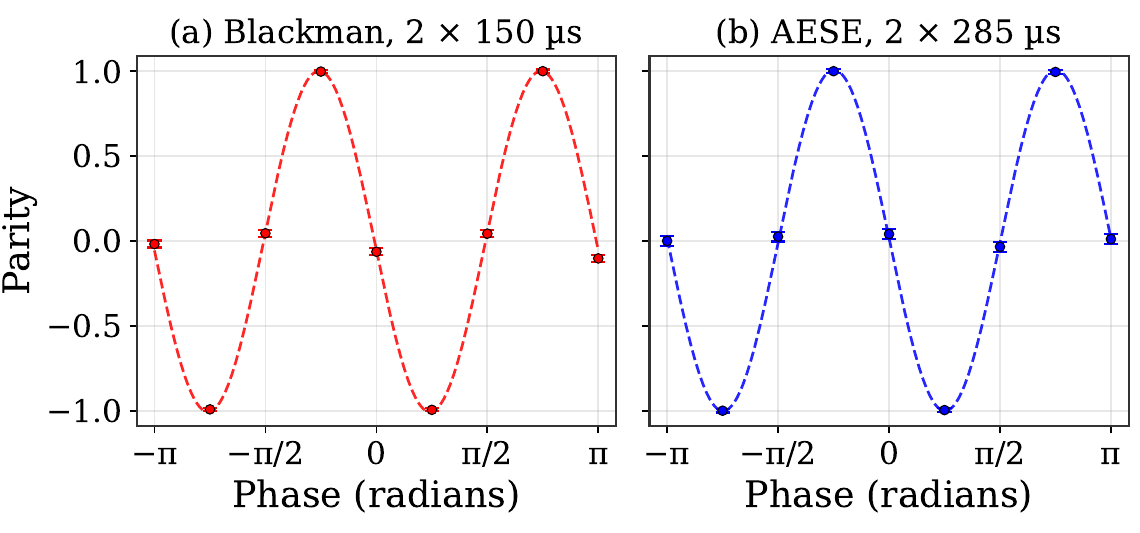}
  \caption[]{Parity $\Pi = P_{uu} + P_{dd} - P_{ud} - P_{du}$ of the qubits after applying the Blackman ramp (a) and AESE (b) sequences illustrated in Figure~\ref{fig:Pulsesequence}, yielding fitted parity contrasts of $1.003(7)$ and $1.004(7)$ and corresponding Bell-state fidelities of $100.0^{+0.0}_{-0.4}\%$ and $100.0^{+0.0}_{-0.3}\%$. Error bars are within the markers, and each point is the result of 4000 (Blackman) or 1900 (AESE) shots. The analysis phase is stepped by $\pi/4$, which samples the fringe at its extremes and zero crossings~\cite{supp}.}
  \label{fig:results}
\end{figure}
\section{Results and Outlook}
\label{sec:results}
\begin{figure}[ht]
\includegraphics[width=\columnwidth]{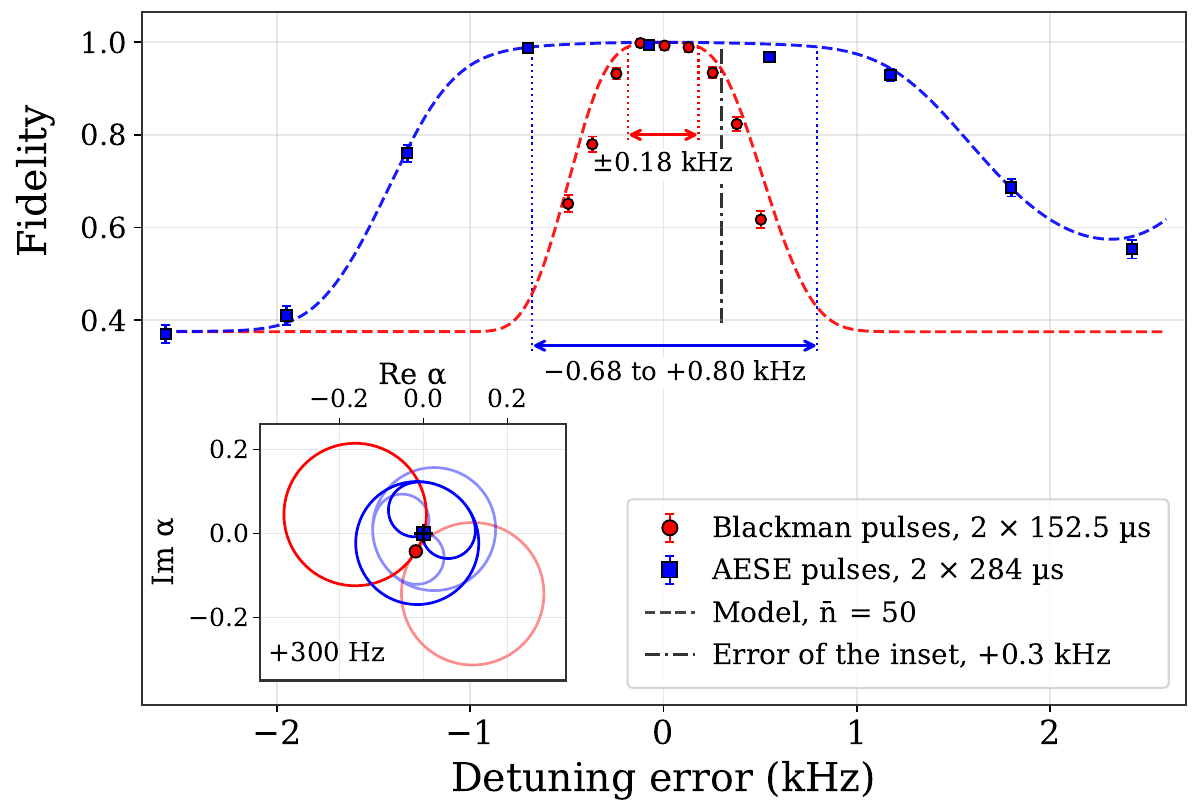}
\caption{Fidelity $F=1-P_{\rm odd}$, where $P_{\rm odd}$ is the population of the odd-parity outcomes after the gate sequence with a $\pi/4$ pulse in place of the $\pi/2$ remapping pulse and no parity pulse, for both the Blackman and AESE cases. Each point is the result of 1000 shots. The fit lines are for a model with a thermal mode occupation of $\bar n=50$, in which only the frequency of the mode is fitted. Red and blue vertical dashed lines mark the range where the fidelity crosses 99\%. The inset shows the paths of the stretch mode for both halves of each sequence in phase space for an error of \SI{300}{\hertz} (dash-dotted line on the main plot). The second half of each sequence is denoted by the fainter lines. The cross marks the origin, and the circle (square) represents the end of the trajectory for the Blackman (AESE) sequence.}
\label{fig:robustness}
\end{figure}

The results of experiments utilising the Blackman and AESE sequences are shown in Figure~\ref{fig:results}. For the Blackman sequence we obtain an unbounded parity contrast fit to the data of $C = 1.003(7)$ and a Bell-state fidelity of $F = 100.0^{+0.0}_{-0.4}\%$, and $C_{AESE} = 1.004(7)$ and $F_{AESE} = 100.0^{+0.0}_{-0.3}\%$ for the AESE implementation~\cite{supp}. Figure~\ref{fig:robustness} also compares the two sequences for varying detuning errors, highlighting the increased robustness of the AESE technique. In this plot the data are fitted to an analytical model of the phase-space trajectories based on Equation~\ref{eq:alpha_general} (see supplementary material). The plot shows the fidelity remains above 99\% for errors between $-0.68$ and \SI{+0.80}{\kilo\hertz} for the AESE pulses, approximately four times as wide as that of the Blackman pulses ($\pm\SI{0.18}{\kilo\hertz}$). The inset shows that for a detuning error of \SI{300}{\hertz}, the Blackman-shaped pulse leaves the stretch mode displaced resulting in a fidelity of 94\%, whereas the AESE sequence returns the mode close to the origin with the fidelity remaining above 99.9\%.
\par
Due to the magnitude of the AC Zeeman shifts specific to our experimental setup, the maximum gradient that can be applied is limited. Using larger gradients produces a stronger out-of-chip field which both begins to drive population out of the qubit basis, and increases the dephasing caused by differential amplitude jitter of the two gate pulses. Simulations of the chip and mounting structure show that the magnitude of these fields can be reduced by a factor of $\sim\!\!100$ if the copper mount were to be replaced with a non-conductive ceramic. Alternatively, a silicon spacer between the chip and the mount would elicit the same effect, with both routes leaving only the much smaller eddy current induced by the metallic chip bond as a limitation, which our simulations show to be negligible. In this case, simulations based on the error model detailed below predict gates on the order of $\SI{90}{\micro\s}$ with fidelities above 99.99\%, for a trap with an ion-height of \SI{75}{\micro\meter} and a gradient of \SI{190}{\tesla\per\meter}. The error budget for the Blackman-ramped gate is shown in Table~\ref{tab:two_qubit_gate_errors}. Further information and derivations regarding these error sources and their values for the AESE gate can be found in the supplementary material.
\par
In summary, we have demonstrated two entangling gates, generating maximally entangled states with fidelities significantly above the relevant fault-tolerant threshold ($F_B=100.0^{+0.0}_{-0.4}\%$ and $F_{AESE}=100.0^{+0.0}_{-0.3}\%$), performed on the axial stretch mode at a mean occupation of $\bar{n}=50(5)$, approximately $2.5\,\times$ the Doppler-cooling limit. Despite significant limitations imposed by hardware (that can be remedied using simple modifications to our setup), we report these fidelities at gate durations comparable to current state-of-the-art laser-free implementations, and realised using only two control signals.
\begin{table}[ht]
\centering
\resizebox{\columnwidth}{!}{%
    \begin{tabular}{lc}
    \hline
    \textbf{Error source} & \textbf{Estimated contribution} \\
    \hline
    Motional errors of the STR mode: & \\
    \quad Heating ($\dot{n}<2$~s$^{-1}$) & $<1.4\times10^{-4}$ \\
    \quad Anharmonicity & $7\times10^{-5}$ \\
    \quad Motional dephasing & $2\times10^{-5}$ \\
    \quad Quasi-static detuning error & $1\times10^{-5}$ \\
    \quad Field inhomogeneity & $2\times10^{-12}$ \\
    Magnetic dephasing & $7(2)\times10^{-5}$ \\
    Differential amplitude jitter of ACG pulses & $5.8\times10^{-5}$ \\
    Residual spin-motion coupling & \\
    (Radial modes through 3rd harmonic of ACG) & $2\times10^{-5}$ \\
    Population leakage out of the subspace & $4\times10^{-8}$ \\
    \hline
    \textbf{Total} & $4\times10^{-4}$ \\
    \hline
    \end{tabular}
}
\caption{Error budget for the Blackman gate~\cite{supp}. The quasi-static detuning error is for a recalibration of the mode frequency every 20~min, and the total takes the heating at its upper bound.}
\label{tab:two_qubit_gate_errors}
\end{table}
\par
A scheme such as this has multiple benefits which make it an attractive option for utility-scale quantum computing. Firstly, removing the need for sub-Doppler cooling significantly simplifies and expedites a quantum computer's clock cycle as it mitigates the need for ground-state cooling. Whilst previous demonstrations of entangling gates using oscillating gradients have achieved very high fidelities using only Doppler cooled ions~\cite{Hughes2025}, they were performed on the radial mode of motion, which cannot be effectively cooled sympathetically using ions of a different species due to the significant mass difference. Because the gate reported here is carried out on the axial mode of motion, the mass difference between ions does not impact the efficiency of sympathetic cooling, therefore enabling the use of global cooling beams or light sheets, significantly relaxing the resource requirements for utility-scale quantum computing. Furthermore, the use of the axial mode is also advantageous for mixed-species entangling gates.
\par
The ability to perform gates well above the Doppler limit is also beneficial for a shuttling-based architecture~\textemdash~excitation-free transport requires careful waveform tuning that is prone to drift induced by stray-fields and trap frequency fluctuations. The ability to tolerate large residual motional excitation increases robustness to zone-to-zone variations, reducing calibration requirements and making the operation of a utility-scale quantum computer robust to small electric field fluctuations and charge up of the chip. In the same vein, gates performed above the Doppler limit are also more tolerant to idle time during algorithms, relaxing circuit scheduling. Additionally, the ability to perform hotter gates eases the constraints on ion-electrode distance imposed by anomalous heating (which scale as $d^{-4}$ with electrode distance $d$), relaxing cryogenic temperature and surface preparation requirements, and permitting smaller $d$ where the gradients driving these gates are stronger. Finally, as the gate only necessitates a single slowly varying oscillating gradient field and resonant microwave pulses to implement, calibration requirements are significantly reduced compared to gates that require a larger number of fields~\cite{Weidt2016}.
\section{acknowledgments}
The authors thank Falk Bonus, Kai-Niklas Schymik, and Juan Diego Arias Espinoza for reading of the manuscript and helpful discussions. The authors acknowledge and thank Arash Pouryazdan, Andrew Betts and Joe Swainston for their technical support. This work was supported by the U.S. Army Research Office under Contract No. W911NF261A198, Contract No. W911NF2110240, Contract No. W911NF1220072, the EPSRC Quantum Information Science \& Technologies Centre for Doctoral Training (EP/Y011864/1), the University of Sussex, Universal Quantum, and the EPSRC QCI3 Hub (EP/Z53318X/1). The first author discloses their responsible use of AI tools for aiding in the development of the models and written analysis detailed in the supplementary material. Any output was analysed and verified by the first author and co-authors, who therefore take full responsibility for any of the tool's results subsequently included in this work.

\bibliography{references}
\clearpage
\input{supplement}

\end{document}

%% file: supplement.tex
\onecolumngrid
\begin{center}
\textbf{\large Supplemental Material: All-electronic geometric phase gate on Doppler-cooled trapped ions using an axial mode}
\end{center}
\vspace{2ex}
\twocolumngrid
\setcounter{secnumdepth}{3}
\setcounter{section}{0}
\setcounter{equation}{0}
\setcounter{figure}{0}
\setcounter{table}{0}
\renewcommand{\thesection}{S\arabic{section}}
\renewcommand{\theequation}{S\arabic{equation}}
\renewcommand{\thefigure}{S\arabic{figure}}
\renewcommand{\thetable}{S\arabic{table}}
\renewcommand{\theHsection}{S\arabic{section}}
\renewcommand{\theHequation}{S\arabic{equation}}
\renewcommand{\theHfigure}{S\arabic{figure}}
\renewcommand{\theHtable}{S\arabic{table}}

This Supplemental Material describes the readout model behind the two-ion populations, its calibration and its
validation (Sec.~\ref{app:spam}), and gives the origin of the gate parameters quoted in the main text: the gradient and
the gate Rabi frequency, the occupation of the stretch mode and the AC Zeeman shift of the gradient pulses
(Sec.~\ref{app:params}). Sections~\ref{app:motion}--\ref{app:leakage} follow the rows of the error budget, Table~I of
the main text: the errors that the motion of the stretch mode causes, with its stability
(Sec.~\ref{app:motion}), magnetic dephasing (Sec.~\ref{app:dephasing}), the amplitude jitter of the gradient pulses
(Sec.~\ref{app:jitter}), the third harmonic of the gradient's drive frequency (the tone) on the radial modes
(Sec.~\ref{app:radial}) and leakage out
of the qubit (Sec.~\ref{app:leakage}). Section~\ref{app:aese} describes the pulses of the gate with adiabatic
elimination of spin--motion entanglement (AESE), the model behind the
comparison of the two sequences in Fig.~3 of the main text, and the contrast measured with the AESE pulses. Figures and tables of the main text are cited by their
numbers there.

\section{State detection and SPAM characterisation}
\label{app:spam}

With $\lambda\approx9$ detected photo-electrons per bright ion, the bright and dark signal distributions of our
camera readout overlap: 16--17\% of the bright ions are pumped dark during the detection window, and the best
threshold misassigns 5.6--5.7\% of the shots of each ion (Fig.~\ref{fig:spam}). We therefore do not use thresholding. The
two-ion populations are maximum-likelihood estimates from the full single-shot signals, with a model of the detection
process that runs from the photon statistics of the ions to the digitised camera output. The model is calibrated on
state-preparation-and-measurement (SPAM) data, and the parameters that can drift are re-fitted from the gate data.

\subsection{Detection}
\label{app:spam:det}

The qubit is read out by state-dependent fluorescence on the $S_{1/2}\,F{=}1\leftrightarrow P_{1/2}\,F'{=}0$
transition at 369.5~nm, with 935-nm repumping~\cite{Olmschenk2007}: an ion in $F{=}1$ (bright, B) scatters photons,
and an ion in $|F{=}0,m_F{=}0\rangle$ (dark, D) does not. These are the states $|{\uparrow}\rangle$ and
$|{\downarrow}\rangle$ of the main text, onto which the gate qubit is mapped before detection. The fluorescence of a 1.0-ms window is imaged onto an
electron-multiplying CCD (EMCCD; Andor iXon Ultra 897, EM gain 200). Each ion is imaged onto a single $4\times4$-binned
super-pixel, so that read noise and digitisation enter once per ion, and a few percent of its light reaches the other
ion's super-pixel. A Sinara Grabber~\cite{Grabber} sums each region of interest (ROI) in real time, so that a shot
gives two integers: the signals $x_1$ and $x_2$, in analogue-to-digital units (ADU), of the ROIs of ion~1 and ion~2.

\subsection{Readout model}
\label{app:spam:model}

\emph{Ions.}---During detection each ion can be represented as a two-state Markov chain~\cite{Acton2006,Wolk2015}. A bright ion is pumped
dark by off-resonant excitation of $P_{1/2}\,F'{=}1$, detuned by $\Delta_P=2.105$~GHz, and a dark ion is pumped bright
by off-resonant excitation from $F{=}0$, detuned by $\Delta_D=14.748$~GHz. While bright, the ion yields detected
photo-electrons as a Poisson process with mean $\lambda$ per window. For an ion that starts the window in state
$s\in\{\mathrm{B},\mathrm{D}\}$, the number $n$ of photo-electrons has the probability generating function (PGF)
$\mathbb{E}[z^n|s]=\mathcal{G}_s(\lambda(z-1))$~\cite{Fischer1993}, with
\begin{equation}
\mathcal{G}_s(u)=\mathbf{e}_s^{\mathsf T}\exp\!\begin{pmatrix}u-\alpha_{\rm bd} & \alpha_{\rm bd}\\
\alpha_{\rm db} & -\alpha_{\rm db}\end{pmatrix}\mathbf{1}.
\label{eq:spam:mmpp}
\end{equation}
Here $\alpha_{\rm bd}$ and $\alpha_{\rm db}$ are the bright$\to$dark and dark$\to$bright pumping rates times the window
length (the pumping rates), $\mathbf{e}_{\rm B}=(1,0)^{\mathsf T}$, $\mathbf{e}_{\rm D}=(0,1)^{\mathsf T}$ and $\mathbf{1}=(1,1)^{\mathsf T}$;
Eq.~\eqref{eq:spam:mmpp} includes any number of state changes within the window. Both pumping processes are off-resonant
excitations by the same light, so their ratio is set by the hyperfine structure,
$\alpha_{\rm db}/\alpha_{\rm bd}=3(\Delta_P/\Delta_D)^2=0.0611$~\cite{Acton2006}, which leaves one pumping parameter per
ion.

\emph{Two ions.}---Each ROI, labelled $r=1,2$ after its ion, also collects a Poisson background of mean $b_r$ electrons (scattered light and spurious
charge) and a fraction $\kappa$ of the other ion's light. For the joint state $\mathbf{s}=(s_1,s_2)$, the joint PGF of
the numbers of electrons $(n_1,n_2)$ that enter the multiplication register from the two ROIs is
\begin{align}
\mathcal{F}_{\mathbf s}(z_1,z_2)={}&e^{b_1(z_1-1)+b_2(z_2-1)}\nonumber\\
&\times\mathcal{G}^{(1)}_{s_1}\!\big(\lambda_1[(z_1-1)+\kappa_{12}(z_2-1)]\big)\nonumber\\
&\times\mathcal{G}^{(2)}_{s_2}\!\big(\lambda_2[\kappa_{21}(z_1-1)+(z_2-1)]\big),
\label{eq:spam:joint}
\end{align}
where $\mathcal{G}^{(i)}$ carries ion $i$'s pumping rates, $\lambda_i$ is its brightness in its own ROI, and
$\kappa_{12}$ ($\kappa_{21}$) is the fraction of that light which ion~1 (2) puts into ROI~2 (1). Crosstalk correlates
the two ROI signals.

\emph{Camera.}---Each electron that enters the $N=536$-stage multiplication register starts a branching process in
which, at each stage, every electron produces a second one with probability $p$, with
$(1+p)^N=G=200$, the EM gain~\cite{Basden2003,Robbins2003,Hirsch2013}. The PGF of the output of one input electron after $k$ stages
follows exactly from $h_k(u)=h_{k-1}(u)\,[1-p+p\,h_{k-1}(u)]$, with $h_0(u)=u$. Electrons created inside the register
by serial clock-induced charge (CIC), $c_{\rm CIC}$ per read at a uniformly distributed stage, are amplified only by the stages
that remain~\cite{Daigle2009,Tulloch2011,Harpsoe2012}. With a conversion of $g$~ADU per input electron, Gaussian read
noise of standard deviation $\sigma$, a bias $o_r$ and rounding to an integer, the characteristic function of the two
ROI signals $\mathbf{x}=(x_1,x_2)$ is
\begin{equation}
\Phi_{\mathbf s}(\boldsymbol\omega)=A_1(\omega_1)\,A_2(\omega_2)\,\mathcal{F}_{\mathbf s}\big(\psi(\omega_1),\psi(\omega_2)\big),
\label{eq:spam:cf}
\end{equation}
where
\begin{align}
A_r(\omega)&=\mathrm{sinc}(\omega/2)\,
e^{\,i\omega o_r-\sigma^2\omega^2/2+c_{\rm CIC}[\psi_{\rm CIC}(\omega)-1]},
\label{eq:spam:camera}\\
\psi(\omega)&=h_N\big(e^{i\omega g/G}\big),\quad
\psi_{\rm CIC}(\omega)=\frac{1}{N}\sum_{k=0}^{N-1}h_k\big(e^{i\omega g/G}\big).\nonumber
\end{align}
Here $\psi$ and $\psi_{\rm CIC}$ are the characteristic functions, in ADU, of the output of a full-gain and of a
serial-CIC electron, and $\mathrm{sinc}\,y=\sin y/y$ accounts for the rounding. The joint probability distribution
of the integer signals,
\begin{equation}
P(\mathbf x\,|\,\mathbf s)=\int_{[-\pi,\pi]^2}\frac{d^2\omega}{4\pi^2}\,
e^{-i\boldsymbol\omega\cdot\mathbf x}\,\Phi_{\mathbf s}(\boldsymbol\omega),
\label{eq:spam:pmf}
\end{equation}
is evaluated by a two-dimensional fast Fourier transform on the integer ADU grid. Because $\sigma\gg1$~ADU,
$\Phi_{\mathbf s}$ vanishes at $|\omega|=\pi$, and on a grid of about 2500~ADU per axis the inversion is exact to
$10^{-5}$.

\emph{Start-of-window failures.}---With probability $\epsilon^{\rm B}_i$ a bright ion behaves as dark from the very
start of the window, and with probability $\epsilon^{\rm D}_i$ a dark ion behaves as bright. Such a shot is identical
to one of the other state, so these probabilities can only be measured on SPAM data. Our preparation errors are much
smaller, about $10^{-6}$ for the dark state and $\lesssim3\times10^{-4}$ for the bright one, so the failures belong to
the detection. They are not correlated between the two ions: in two-ion SPAM runs with $2.2\times10^5$ shots with both ions
bright, both failed together in less than 0.018\% of the shots (95\% confidence). Nor do they come from the microwave
$\pi$ pulse of the bright preparation, which acts in the same way on both ions: the failure probability depends on
where the ion sits, 0.23(2)\% at the position of ion~1 and 0.51(12)\% at that of ion~2 in single-ion runs taken
alternately. We attribute the failures to the timing of the camera. Its exposure output shows that the
exposure starts between 0.34 and 0.52~ms after the camera is triggered, while the detection light was switched on
0.2~ms after the trigger. The sensor keeps the light that arrives before the exposure, but shifts it by one row in
each of its cleaning cycles, which last 0.18~ms, so that part of it leaves the ion's ROI. A bright ion that is
pumped dark in this time has given its light before the exposure, and where that light has left the ROI the shot is
that of a dark ion. How much leaves depends on where the image of the ion lies in its ROI, which is the dependence
on the position. In a separate experiment, the light was switched on 0.4~ms after the trigger instead: in 13
single-ion runs with $1.25\times10^6$ shots per state the failure probability of a bright ion was $0.09(2)\%$,
against $0.53(7)\%$ in three runs with 0.2~ms before the new setting was applied, and a photomultiplier,
which has no such timing, gives $0.10\%$ for the same preparation. The likelihood of a shot for the joint state $\mathbf S$ of the ions before detection is therefore
\begin{align}
L(\mathbf x\,|\,\mathbf S)&=\sum_{\mathbf s}(K_1\otimes K_2)_{\mathbf s\mathbf S}\,P(\mathbf x\,|\,\mathbf s),
\label{eq:spam:failures}\\
K_i&=\begin{pmatrix}1-\epsilon^{\rm D}_i & \epsilon^{\rm B}_i\\ \epsilon^{\rm D}_i & 1-\epsilon^{\rm B}_i
\end{pmatrix},\nonumber
\end{align}
with the rows and columns of $K_i$ in the order (D, B).

\begin{figure}[t]
\includegraphics[width=\columnwidth]{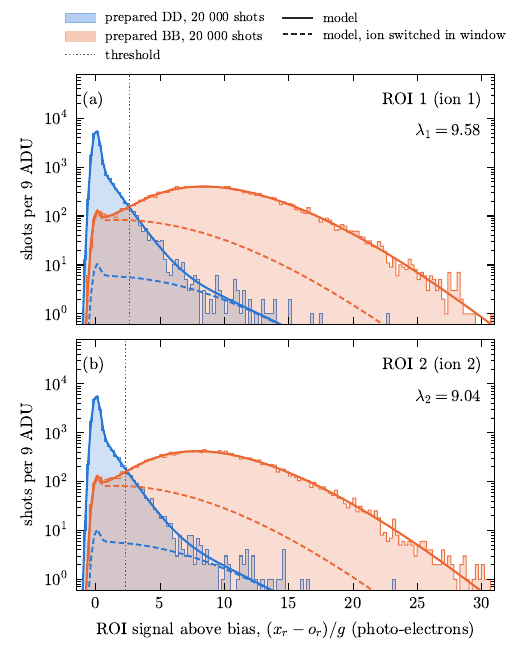}
\caption{SPAM characterisation of the two-ion readout, the two preparations interleaved shot by shot. The model has
the parameters of Table~\ref{tab:spam} and no free normalisation; it fits the counts down to their shot noise (reduced $\chi^2$ of 0.84--1.05). The ROI's own ion changed state at least once during the 1.0-ms window in 17.2\% (16.1\%)
of the bright and 1.2\% (1.1\%) of the dark shots of ion~1 (2); these shots fill the region between the peaks.
The thresholds give fidelities of 94.4\% (ROI~1) and 94.3\% (ROI~2).}
\label{fig:spam}
\end{figure}

\begin{table}[t]
\caption{The parameters of the readout model for the SPAM run of Fig.~\ref{fig:spam}, and how the gate analysis treats
them: re-fitted from every gate scan with a weak prior around the calibration value, or held at the calibration
value. The 13 parameters of the detection model are fitted to the 40\,000 shots of the run, with errors from the
Hessian of the likelihood; the failure probabilities come from single-ion SPAM runs and are held in this fit; the scans of
Sec.~\ref{app:spam:gate} use a value for each ion (Sec.~\ref{app:spam:cal}), and re-fit ion~2's pumping rate, for which
they have no calibration. The
dark$\to$bright pumping rates are tied, $\alpha_{{\rm db},i}=0.0611\,\alpha_{{\rm bd},i}$.}
\label{tab:spam}
\begin{ruledtabular}
\begin{tabular}{lccc}
 & ion/ROI 1 & ion/ROI 2 & gate \\
\hline
Bias $o_r$ (ADU) & 486.6(1) & 499.0(1) & re-fitted \\
Background $b_r$ ($e^-$) & 0.351(6) & 0.325(6) & re-fitted \\
Gain $g$ (ADU$/e^-$) & \multicolumn{2}{c}{37.83(9)} & re-fitted \\
Serial CIC $c_{\rm CIC}$ ($e^-$/read) & \multicolumn{2}{c}{0.18(3)} & re-fitted \\
Read noise $\sigma$ (ADU) & \multicolumn{2}{c}{8.77(6)} & held \\
Brightness $\lambda_i$ ($e^-$) & 9.58(8) & 9.04(8) & re-fitted \\
Crosstalk $\kappa_{12}$, $\kappa_{21}$ (\%) & 4(1) & 4(1) & re-fitted \\
Leak B$\to$D $\alpha_{{\rm bd},i}$ & 0.189(9) & 0.176(8) & held \\
Failure B $\epsilon^{\rm B}_i$ (\%) & \multicolumn{2}{c}{0.26(4)} & held \\
Failure D $\epsilon^{\rm D}_i$ (\%) & \multicolumn{2}{c}{0.011(6)} & held \\
\end{tabular}
\end{ruledtabular}
\end{table}

\begin{table}[t]
\caption{Model selection on the SPAM run of Fig.~\ref{fig:spam} (40\,000 shots). Each row removes one ingredient
from the detection model of Table~\ref{tab:spam} ($k=13$ parameters), or unties the dark$\to$bright pumping rates, and refits
with the failure probabilities held:
$\Delta k$ is the change in the number of parameters, $2\Delta\ln L$ the change in twice the maximum log-likelihood,
$p$ the likelihood-ratio test, and $\Delta\mathrm{AIC}=2\Delta k-2\Delta\ln L$ the change in the Akaike information
criterion, which is positive where the model of Table~\ref{tab:spam} is preferred.}
\label{tab:dof}
\begin{ruledtabular}
\begin{tabular}{lrrrr}
Model & $\Delta k$ & $2\Delta\ln L$ & $p$ & $\Delta\mathrm{AIC}$ \\
\hline
No pumping (Poisson ions) & $-2$ & $-1358$ & $<10^{-200}$ & $+1354$ \\
No background & $-2$ & $-1143$ & $<10^{-200}$ & $+1139$ \\
No crosstalk & $-2$ & $-34.9$ & $3\times10^{-8}$ & $+30.9$ \\
No serial CIC & $-1$ & $-7.1$ & 0.008 & $+5.1$ \\
Dark$\to$bright pumping untied & $+2$ & $+1.1$ & 0.57 & $+2.9$ \\
\end{tabular}
\end{ruledtabular}
\end{table}

\subsection{Calibration}
\label{app:spam:cal}

Figure~\ref{fig:spam} shows an example two-ion SPAM run of another crystal (a crystal is one loaded pair of ions): 20\,000 shots with both ions prepared dark (DD) and 20\,000 with both prepared bright (BB),
interleaved shot by shot. Each shot consists of Doppler cooling, 250~$\mu$s of optical pumping into
$|F{=}0,m_F{=}0\rangle$, for BB a microwave $\pi$ pulse to $|F{=}1,m_F{=}{+}1\rangle$, and the 1.0-ms detection.
The 13 parameters of the detection model, five for each ion (its brightness and pumping rate, and the background,
bias and crosstalk fraction of its ROI) and three for the camera (gain, serial CIC and read noise), are fitted together by maximising the likelihood of all 40\,000 shots under the exact joint distribution,
Eq.~\eqref{eq:spam:pmf}. The camera parameters are thus those of the camera during the run, and the crosstalk is
determined by the correlation between the two ROI signals.

The model describes the data with no free normalisation (Fig.~\ref{fig:spam}). For the four histograms of
Fig.~\ref{fig:spam} the reduced $\chi^2$ is 0.84--1.05: the model fits the counts down to their shot noise. Across
all six SPAM runs of that day, with one or two ions and $10^4$--$10^5$ shots per state, the 24 histograms fit as well
as counts drawn from the model itself.

The model has as many parameters as the data support (Table~\ref{tab:dof}). Removing any ingredient --- the pumping, the
background, the crosstalk or the serial CIC --- makes the fit significantly worse ($p\leq0.008$), while untying the
dark$\to$bright pumping rates does not improve it ($p=0.57$). A shutter-closed run of the
same evening measures the serial CIC independently, $c_{\rm CIC}=0.15(4)$. An ion is pumped dark after about
$\Lambda=\lambda/\alpha_{\rm bd}\approx50$ detected photo-electrons; these events, not the camera noise, cause most of
the overlap between the bright and dark distributions, which is an inherent limit of hyperfine detection with
$^{171}$Yb$^+$.

Single-ion runs, with many more shots per ion, measure the failure probabilities better than a two-ion run. From 16
single-ion SPAM runs with $10^5$ shots per state, $\epsilon^{\rm B}=0.26(4)\%$ and $\epsilon^{\rm D}=0.011(6)\%$, which
we hold for both ions in the fit of Fig.~\ref{fig:spam}; the run of Fig.~\ref{fig:spam} agrees with them
($2\Delta\ln L=4.4$ for the four held, $p=0.35$).

The two parity scans of the main text (Sec.~\ref{app:spam:gate})
were calibrated on a single-ion SPAM run of the same day, taken 3.5~h before them and before
their crystal was loaded, with $10^5$ shots per state, fitted in the same way. In the fit of a scan, ion~2's
parameters start from ion~1's values.

In these scans each ion has its own failure probability, since the light that leaves a ROI before the
exposure depends on where the image of the ion lies in it (Sec.~\ref{app:spam:model}). For ion~1 it is that of the
single-ion run of the same day, whose ion sat in ROI~1: $\epsilon^{\rm B}_1=0.22(11)\%$. For ion~2 it is inferred
from the SPAM runs that have an ion in ROI~2 (Table~\ref{tab:data}): $\epsilon^{\rm B}_2=0.46(17)\%$, with an
uncertainty that covers the spread between these runs. The same light pumps a dark ion bright 0.0611 times as often,
so that $\epsilon^{\rm D}_2=0.028(10)\%$, and $\epsilon^{\rm D}_1$ is zero within its error. An error of 0.1\% in
$\epsilon^{\rm B}_2$ changes the contrast of Sec.~\ref{app:spam:gate} by 0.001.
Table~\ref{tab:data} lists the data that enter the gate analysis.

\begin{table}[t]
\caption{The data behind the gate analysis of Sec.~\ref{app:spam:gate} and what each provides. Days are counted
from the parity scans of the main text.}
\label{tab:data}
\def\datacell#1#2{\parbox[t]{#1\columnwidth}{\raggedright\strut#2\strut}}%
\begin{ruledtabular}
\begin{tabular}{ll}
Data & Use in the gate analysis \\
\hline
\datacell{0.36}{Single-ion SPAM run, same day, 3.5~h before ($10^5$ shots per state)} &
\datacell{0.56}{Held: read noise $\sigma$, ion~1's pumping rate $\alpha_{{\rm bd},1}$ and ion~1's failure
probabilities. Starting values and prior
centres of the parameters re-fitted per scan, ion~2's from ion~1's} \\
\datacell{0.36}{The parity scans of the main text} &
\datacell{0.56}{Per scan: $\lambda_{1,2}$, $b_{1,2}$, $\kappa_{12,21}$, $g$, $c_{\rm CIC}$, $o_{1,2}$ and ion~2's pumping rate} \\
\datacell{0.36}{16 single-ion SPAM runs, 2--3 days after ($10^5$ shots per state each)} &
\datacell{0.56}{Ion~2's failure probabilities (ion centred in ROI~2); the common values held in the fit of
Fig.~\ref{fig:spam}} \\
\datacell{0.36}{Two-ion SPAM run, same day, another crystal ($10^5$ shots per state)} &
\datacell{0.56}{Ion~2's failure probabilities (ion in ROI~2)} \\
\datacell{0.36}{Two-ion SPAM runs, 3--4 days after ($2.2\times10^5$ shots)} &
\datacell{0.56}{That the two ions fail independently (the form $K_1\otimes K_2$); no number is taken from
them} \\
\datacell{0.36}{The two-ion SPAM run of Fig.~\ref{fig:spam}, 5 days before, another crystal} &
\datacell{0.56}{Table~\ref{tab:spam}, ion~2's failure probabilities (ion in ROI~2), and the test of the
one-ion calibration on that crystal's own gate scans (Sec.~\ref{app:spam:gate})} \\
\end{tabular}
\end{ruledtabular}
\end{table}

\subsection{Gate populations and parity contrast}
\label{app:spam:gate}

The parity scans of the main text are two consecutive scans of
the gate of Table~\ref{tab:params} on one crystal, each taken directly after a measurement of the stretch-mode
frequency (Sec.~\ref{app:motion}). A parity scan runs the sequence of Fig.~1 of the main text at 9 phases
$\phi_m$ of its parity pulse, here with 2000 shots each. The phases run from $-\pi$ to $\pi$ in steps of $\pi/4$ and sample the
fringe, whose period is $\pi$, at its extremes and at its zero crossings, each twice: the extremes fix the contrast
and the zero crossings the phase. Neither they
nor the run of Fig.~\ref{fig:spam} show ion drop-outs (runs of consecutive dark shots).
For each scan, the populations at every phase and the readout parameters that drift between calibration and scan are
fitted together by maximising
\begin{equation}
\ln\mathcal{L}=\sum_{m,j}\ln\sum_{\mathbf s}\pi^{\rm seen}_{m,\mathbf s}\,
P(\mathbf x_{mj}\,|\,\mathbf s;\theta)+\ln p(\theta),
\label{eq:spam:gatelik}
\end{equation}
where $\mathbf x_{mj}$ are the two signals of shot $j$ at phase $\phi_m$ and $\pi^{\rm seen}_m$ are the four
populations at that phase as the detection sees them, three free parameters per phase, with no fringe shape assumed. The vector $\theta$ holds the parameters that Table~\ref{tab:spam} marks as re-fitted and
ion~2's pumping rate, which the single-ion calibration does not give; $p(\theta)$ is a weak prior around the calibration
values. A $9\times2000$ scan thus has $11+9\times3=38$ free parameters for 18\,000 shots. Every phase mixes the same four
distributions with different weights, so the scan itself determines them. In simulated scans, re-fitting ion~2's pumping
rate biases $C$ by $-0.3\%$ to $+0.2\%$, depending on its true value. The read noise, ion~1's pumping rate and the
failure probabilities are held at their calibration values.

The populations are the maximum-likelihood estimates without a positivity constraint, which would bias the near-zero
populations at the fringe extremes, and with them the contrast; a point near an extreme can therefore lie beyond
$\pm1$ by up to its error. Figure~2 of the main text shows each point with the populations held
non-negative, in panel~(a) from the shots of the two scans at that phase together, while the contrasts are fitted to
the unconstrained points. The covariance of the populations follows from the observed Fisher information, so it contains
both projection noise and the ambiguity of the readout. The populations before detection are
$\pi_m=(K_1\otimes K_2)^{-1}\pi^{\rm seen}_m$.

The parity $\Pi_m=\pi_{m,\rm DD}+\pi_{m,\rm BB}-\pi_{m,\rm DB}-\pi_{m,\rm BD}$ is fitted with $c+C\cos(2\phi_m+\phi_0)$, with the contrast $C$, the offset $c$ and the phase $\phi_0$,
by unweighted least squares on the eight distinct phases (the points at $-\pi$ and $\pi$ are averaged), where the fit
is the exact Fourier component of the parity at $2\phi$. The errors of $C$ and $c$ are
propagated from the covariance of the populations. Shifting each fixed pumping rate and failure probability by its
calibration error gives $\pm0.0024$ on $C$, most of it from ion~2's failure probability.

Calibrating on a single ion adds an uncertainty: ion~1's pumping rate, held at the single ion's value, need
not be that of the crystal's ion~1. We measured the size of this effect on the crystal of Fig.~\ref{fig:spam}, whose
gate scans have both a two-ion calibration of their own and a single-ion run of the same day: taking ion~1's pumping
rate from the single ion, which was pumped 11\% faster than the crystal's ion~1, changed $C$ by 0.004. We add
$\pm0.005$ for the one-ion calibration: the change of $C$ just described and the bias of up to 0.003 that re-fitting
ion~2's pumping rate gives in simulated scans, combined in quadrature. Together with the $\pm0.0024$ above, the
readout error is $\pm0.0055$, common to the two scans.
The scans give $C=1.0038\pm0.0065$ and $1.0030\pm0.0066$; together, with statistical weights,
$C=1.0034\pm0.0046\,(\mathrm{stat})\pm0.0055\,(\mathrm{readout})$ and $c=-0.003\pm0.005$. The main text gives these
numbers with three decimals.

The fidelity to the Bell state $(|\mathrm{DD}\rangle+e^{i\chi}|\mathrm{BB}\rangle)/\sqrt2$, with $\chi$ fixed by the phase $\phi_0$ of the fringe, is $F=(P_{\rm DD}+P_{\rm BB})/2+C/2$, where $P_{\rm DD}$ and $P_{\rm BB}$ are the populations of
$|\mathrm{DD}\rangle$ and $|\mathrm{BB}\rangle$. Positivity of the density matrix bounds the DD--BB coherence, $C/2$, by
$(P_{\rm DD}+P_{\rm BB})/2$, so that $F\geq C$ for any two-ion state, while $F\le1$. The measured contrast
lies at this limit, and a fidelity written as $C$ with a symmetric error would have most of its range above one. We give instead
the confidence interval of Feldman and Cousins for a Gaussian measurement of a quantity with a physical
limit~\cite{Feldman1998}: for every true value $\mu\le1$ the outcomes $C$ are accepted in order of the likelihood
ratio $P(C|\mu)/P(C|\min(C,1))$ until they hold 68\%, and the interval is the set of $\mu$ that accept the measured
$C$. Here $P(C|\mu)$ is a Gaussian about $\mu$ of width
$\sigma_C=[(1.09\,\sigma_{\rm stat})^2+\sigma_{\rm ro}^2]^{1/2}=0.0075$,
where $\sigma_{\rm stat}$ and $\sigma_{\rm ro}$ are the statistical and the readout error of $C$ and the statistical error
is widened by the spread found on synthetic scans (Sec.~\ref{app:spam:val}). The interval runs from 0.9956 to 1,
\begin{equation}
F=100.0^{+0.0}_{-0.4}\%,
\label{eq:spam:fidelity}
\end{equation}
and it is how the main text and its appendices state the fidelity, beside the contrast with its two errors combined
in quadrature, $C=1.003(7)$. The interval is not a standard uncertainty: its lower part is this short because the
contrast came out above one; a perfect gate gives a contrast this high or higher in 32\% of the cases with this
error.

\subsection{Validation}
\label{app:spam:val}

The model reproduces a Monte Carlo simulation of the full signal chain, at the pixel level, with $10^7$ shots per
state, within statistical noise. Gate scans drawn from the model, with $9\times2000$ shots and readout parameters
typical of our runs, were analysed with the full procedure for a crystal with a two-ion calibration, readout
re-fit included (150 scans per case). A cosine
fitted to the populations of each gives $C=0.9994(6)$ for a true contrast of 1 and 0.9804(6) for 0.98, and the quoted
errors match the scatter (1.06 and 1.02 times the quoted error). Over 600 such scans, with 1000 or 2000
shots per phase, the contrast scatters by 1.09 times its quoted statistical error; this is the factor in the error
of the interval of Sec.~\ref{app:spam:gate}.\space\space

\section{Gate parameters}
\label{app:params}

Table~\ref{tab:params} lists the settings of the gate for the Blackman pulse sequence, whose fidelity is given in the main text. This section gives the
origin of three of its entries: the gradient and the gate Rabi frequency, the occupation of the stretch mode, and the
AC Zeeman shift of the gradient pulses.

\begin{table}[tb]
\caption{Settings of the gate (the two parity scans of the main text, on one crystal). The gradient is that of the magnitude of the
field at the tone's frequency, which the qubit feels (Sec.~\ref{app:params:acz}), for the pulses as run,
Eq.~\eqref{eq:par:grad}; the gate Rabi frequency is that of rectangular loops, Eq.~\eqref{eq:par:omega}.}
\label{tab:params}
\begin{ruledtabular}
\begin{tabular}{lr}
Static field & 16.77~G \\
Axial modes, centre-of-mass and stretch & 288.5, 499.63~kHz \\
Radial modes (in phase) & 1.384, 1.464~MHz \\
Trap drive & 15.70~MHz \\
Ion separation & 7.9~$\mu$m \\
Gradient tone & 506.88~kHz \\
Tone above the stretch mode & 7.25~kHz \\
Two gradient pulses, total length & $2\times150~\mu$s \\
Pulse edges (Blackman) & 10~$\mu$s \\
Gap between the pulses & 44~$\mu$s \\
First microwave to remapping pulse & 0.38~ms \\
Gate Rabi frequency $\Omega_g/2\pi$ & 2.56~kHz \\
Gradient & 35(1)~T/m \\
Stretch-mode occupation $\bar n$ & 50(5) \\
AC Zeeman shift of the qubit & 5.46~MHz \\
AC Zeeman phase per pulse $\Phi_{\rm Z}$ & $4.9\times10^3$~rad \\
\end{tabular}
\end{ruledtabular}
\end{table}

\subsection{Gradient and gate Rabi frequency}
\label{app:params:grad}

We use the notation of the main text throughout. In the stretch mode the two ions move in opposite
directions, so the oscillating gradient couples the mode to $\hat S_{z,m}=\sigma_{z,1}-\sigma_{z,2}$, which is
$\pm2$ on the two odd-parity states and zero on the two even-parity states. In the frame that rotates with the qubits
and the mode, and in the rotating-wave approximation,
\begin{align}
\hat H&=\frac{\hbar\Omega_g}{2}\,g(t)\,\hat S_{z,m}\big(\hat a^{\dagger}e^{i\varphi(t)}+\hat a\,e^{-i\varphi(t)}\big),
\label{eq:par:H}\\
\Omega_g&=\frac{1}{2\sqrt2}\,\frac{\partial\omega_0}{\partial B}\,\frac{\partial B}{\partial z}\,z_0 ,
\label{eq:par:Omega}
\end{align}
where $g(t)$ is the pulse envelope. The frequency of the oscillating gradient lies at
$\omega_s-\delta$, and $\varphi(t)=\int_0^t\delta\,dt'$. Our tone lies above the mode, so that $\delta$ is negative, and
$\varphi=\delta t$ for a constant detuning. Further, $\omega_0$ is the qubit frequency, $\partial B/\partial z$ the
gradient along the trap axis $z$ that the qubit feels (Sec.~\ref{app:params:acz}), $z_0=\sqrt{\hbar/2M\omega_s}
=7.69$~nm the extent of the ground state of one ion of mass $M$ at the stretch frequency $\omega_s$, and the
factor $1/\sqrt2$ the share of each ion in the mode. With $\partial\omega_0/\partial B=2\pi\times28.0$~GHz/T,
$\Omega_g/2\pi$ is 76.2~Hz for each T/m of gradient. $\Omega_g$ is the rate at which the force
displaces the odd-parity states, and the value in the table represents the drive required to generate an entangled state for a square pulse without the dispersive effects of both the COM mode (a consequence of our negative detuning), and the counter-rotating drive at $2\omega_s-\delta$. The factor $\tfrac12$ in Eq.~\eqref{eq:par:H} is therefore the co-rotating half of the
oscillating force at $\sim\delta$ i.e. that which is kept when the rotating wave-approximation is made.

A pulse displaces the mode by $\alpha\hat S_{z,m}$ and, once its loop is closed, leaves the phase
$\Phi\hat S_{z,m}^2$, with
\begin{equation}
\alpha(t)=-\frac{i\Omega_g}{2}\!\int_0^t\!g(t')\,e^{i\varphi(t')}dt',\quad \Phi=\mathrm{Im}\!\int\!\alpha^*d\alpha .
\label{eq:par:alpha}
\end{equation}
Since $\hat S_{z,m}^2=2(1-\sigma_{z,1}\sigma_{z,2})$, a geometric phase $|\Phi|=\pi/8$ gives
$\exp(\pm i\tfrac{\pi}{4}\sigma_{z,1}\sigma_{z,2})$ up to a global phase, the maximally entangling gate.
A rectangular pulse that closes one loop of length $2\pi/|\delta|$ contributes
$|\Phi|=\tfrac{\pi}{2}(\Omega_g/\delta)^2$, so that two loops require
\begin{equation}
\Omega_g=\frac{|\delta|}{2\sqrt2}=2\pi\times2.56~\mathrm{kHz}\quad\text{at}\quad\delta=-2\pi\times7.25~\mathrm{kHz}.
\label{eq:par:omega}
\end{equation}

For the pulses as run we integrate the motion numerically. With the Blackman ramps of 10~$\mu$s, the ideal
detuning increases compared to a plain square pulse from $|\delta|/2\pi=6.67$~kHz to 7.226~kHz. Given calibration
imperfections, the Blackman gate data sets used an actual detuning of 7.250~kHz, and consequently the first pulse
leaves a small residual displacement $\alpha_1$ (see Sec.~\ref{app:motion} for the analysis of the resulting
geometric phase and motional dephasing errors).
The counter-rotating term lowers the phase by 0.7\%, and the centre-of-mass mode, which the same tone drives
218~kHz from its resonance, by a further 4\%~\cite{Sutherland_2024}. The peak gate Rabi frequency
must therefore be 2.5\% above the $2\pi\times2.56$~kHz of Eq.~\eqref{eq:par:omega}, $\Omega_g=2\pi\times2.63$~kHz. Given the data are consistent with a maximally entangled state, a magnetic field gradient of
\begin{equation}
\frac{\partial B}{\partial z}=35(1)~\mathrm{T/m}
\label{eq:par:grad}
\end{equation}
is inferred, the uncertainty of which is that of the geometric phase. Since $1-F\geq4\,\delta\Phi^2$
[Eq.~\eqref{eq:mot:err}], the measured fidelity allows an error of the phase of up to 0.032~rad, 8\% of $\pi/8$, which
is 4\% of the gradient.

\subsection{Occupation of the stretch mode}
\label{app:params:nbar}

At the detuning at which its loops close, the gate does not depend on the occupation of the mode, which can be measured via its response to the set detuning. In a detuning scan the sequence of Fig.~1 of the main text is run without the parity
pulse and with a $\pi/4$ pulse in place of the $\pi/2$ remapping pulse (the first microwave pulse after the
second gradient pulse), which maps the output of an ideal gate onto equal
populations of $|{\downarrow\downarrow}\rangle$ and $|{\uparrow\uparrow}\rangle$, while the tone is scanned by several
hundred hertz around the closing detuning. Away from the closing detuning each pulse leaves a displacement $\alpha(\delta)$ on the mode, which entangles the spins with the motion and so moves population into the odd outcomes. That population is $4(\bar n+\tfrac12)|\alpha|^2$ while it remains small, and saturates at $5/8$ once the motion is fully entangled with the spins. Here $\bar n$ sets only how large a displacement is needed to reach this saturation point, not the saturation point itself. The saturated population is $\frac{1}{2}+\frac{1}{4}\sin^2\theta$ for an analysis pulse of angle $\theta$, which is $5/8$ at $\theta=\pi/4$. Since the force is fixed by the geometric phase, $|\alpha(\delta)|$ is the only quantity that varies across the scan, and the width of the region in which the odd population stays small depends on $\bar n$ alone. We fit such scans with the thermal average of the two-loop sequence, with $\alpha(\delta)$ and $\Phi(\delta)$ computed
for the pulses as run, one occupation for all scans of a gate, and an offset of the closing detuning and a constant floor of the odd population
for each scan.
The 13 scans of a gate of two 155-$\mu$s pulses and the 7 scans of a gate of two 150-$\mu$s pulses, taken in an
earlier session, give $\bar n=50(2)$ and $51(3)$, with $\chi^2=277$ and 249 for 244 and 250 degrees
of freedom. The occupation expected at the Doppler limit of the cooling transition, $\bar n\approx20$, is excluded
($\Delta\chi^2>300$), even when a spread of the mode frequency from shot to shot is allowed for. Changing the gap between the pulses by $\pm5~\mu$s and the geometric phase by $\pm10\%$ moves $\bar n$
between 46 and 58. Scans taken in later sessions give $\bar n=52(4)$ with Blackman-edged pulses and
$54(4)$ with the AESE pulses (Sec.~\ref{app:aese}). We
quote $\bar n=50(5)$, a temperature of 1.2~mK, for the occupation during the gate.

\subsection{AC Zeeman shift of the gradient pulses}
\label{app:params:acz}

\emph{Mechanism.}---The ions sit about 30~$\mu$m from the null of the oscillating quadrupole field, where
the gradient comes with a field $b_\perp\cos\omega_gt$ perpendicular to the static field $B_0$, which lies along
the trap axis. The tone's frequency
$\omega_g$ is far below the Larmor frequency, so the spins follow the total field, and the levels
$|F,m_F\rangle=|1,\pm1\rangle$ move
with its magnitude. Averaged over a period of the tone, the qubit frequency shifts by
\begin{align}
\Delta_{\rm q}&=2\gamma\Big(\big\langle\sqrt{B_0^2+b_\perp^2\cos^2\omega_gt}\,\big\rangle-B_0\Big)\nonumber\\
&\simeq\frac{\gamma\,b_\perp^2}{2B_0}\Big(1-\frac{3b_\perp^2}{16B_0^2}\Big),
\label{eq:par:acz}
\end{align}
with the gyromagnetic ratio $\gamma=2\pi\times1.40$~MHz/G of the levels $|1,\pm1\rangle$, each of the two levels by half of it. The shift is of second order in the drive.
The qubit frequency is also modulated at $2\omega_g$, by about as much as it is shifted, which averages out over a
pulse.

The total field also tilts, by up to $\arctan(b_\perp/B_0)$, and the qubit feels the gradient of its magnitude,
$\partial|B|/\partial z\simeq(B_0/|B|)\,\partial B_z/\partial z$. Its component at the tone's frequency is the
gradient of the axial field component times $\langle2\cos^2\!\omega_gt\;B_0/|B|\rangle$, the average of
$B_0/|B|$ over a period of the tone with the weight $2\cos^2\!\omega_gt$, which is below one. The gradient of
Eq.~\eqref{eq:par:grad} and of the main text is this component, the one that drives the gate.

\emph{Measurement.}---We measured the drive field with a single ion, on the transition
$|0,0\rangle\to|1,{+}1\rangle$ with the drive on. Under the drive the line shifts by $\Delta_{\rm q}/2$ and splits into
a comb of lines $2\omega_g$ apart, with weaker lines at odd multiples of $\omega_g$ where the ion is displaced from the
axial null of the field. We recorded 53 frequency scans at a range of drive amplitudes up to that of the
gate, probe
pulses of 4.5 to 447~$\mu$s and twelve axial positions of the ion, and fitted them together with the line shape that
follows from the exact evolution of the $F=1$ manifold over a period of the drive. Scaled to the current in the
wires during the gate, the fit gives a linearly polarised transverse field of $b_\perp=11.9$~G at the ions,
71\% of the static field.

\emph{The gate.}---With this field each of the two levels shifts by 2.73~MHz and the qubit by
$\Delta_{\rm q}=2\pi\times5.46$~MHz, the quantisation axis tilts by up to $35^\circ$, and the weighted average of $B_0/|B|$ is 0.85. Integrated
over a pulse, the shift leaves the phase
\begin{equation}
\Phi_{\rm Z}=\int\!\Delta_{\rm q}(t)\,dt\approx4.9\times10^3~\mathrm{rad}
\label{eq:par:phase}
\end{equation}
on each ion, and about 2\% more on one ion than on the other, since the transverse field changes along the trap axis.
The echo pulse reverses the sign of this phase, and the second gradient pulse removes it as far as the two pulses are
equal (Sec.~\ref{app:jitter}).\space\space

\section{Motional errors}
\label{app:motion}

During each gradient pulse the spins are entangled with the motion of the stretch mode. Sutherland et al.\
list the errors that the motion then causes in gates of this kind~\cite{Sutherland2022}: a static shift of the mode
frequency (a quasi-static detuning error), noise of the mode frequency (motional dephasing), anharmonicity of the
motion, inhomogeneity of the field that exerts the force, and heating. This section estimates each of them for the two sequences of the main text, the pair of Blackman-edged
pulses (Table~\ref{tab:params}) and the pair of AESE pulses (Sec.~\ref{app:aese}), at $\bar n=50$.
Table~\ref{tab:motion} collects the results.

\emph{The trajectory.}---In the frame that rotates with the mode at its calibrated frequency, the force
displaces the two odd-parity states by $\pm2\alpha(t)$, the eigenvalues of $\hat S_{z,m}$ times $\alpha$; during the first
pulse $\alpha$ is that of Eq.~\eqref{eq:par:alpha}. The microwave echo pulse between the two gradient pulses flips both spins and so reverses the
force, and the phase of the second pulse is set such that it continues the first on the mode. The second pulse
therefore removes the displacement left by the first, so that $\alpha$ returns to zero at the end of the gate, and
adds the same geometric phase. The force is
fixed by the geometric phase, $|\Phi|=\pi/8$ (Sec.~\ref{app:params:grad}). A small perturbation leaves a displacement
$\delta\alpha$ at the end of the gate and changes the geometric phase by $\delta\Phi$. For a thermal mode the fidelity of
the Bell state is then
\begin{equation}
1-F\simeq4\big(\bar n+\tfrac12\big)|\delta\alpha|^2+4\,\delta\Phi^2 ,
\label{eq:mot:err}
\end{equation}
as long as both terms are small. Section~\ref{app:aese} gives the expression of which this is the expansion.

\begin{figure}[t]
\includegraphics[width=\columnwidth]{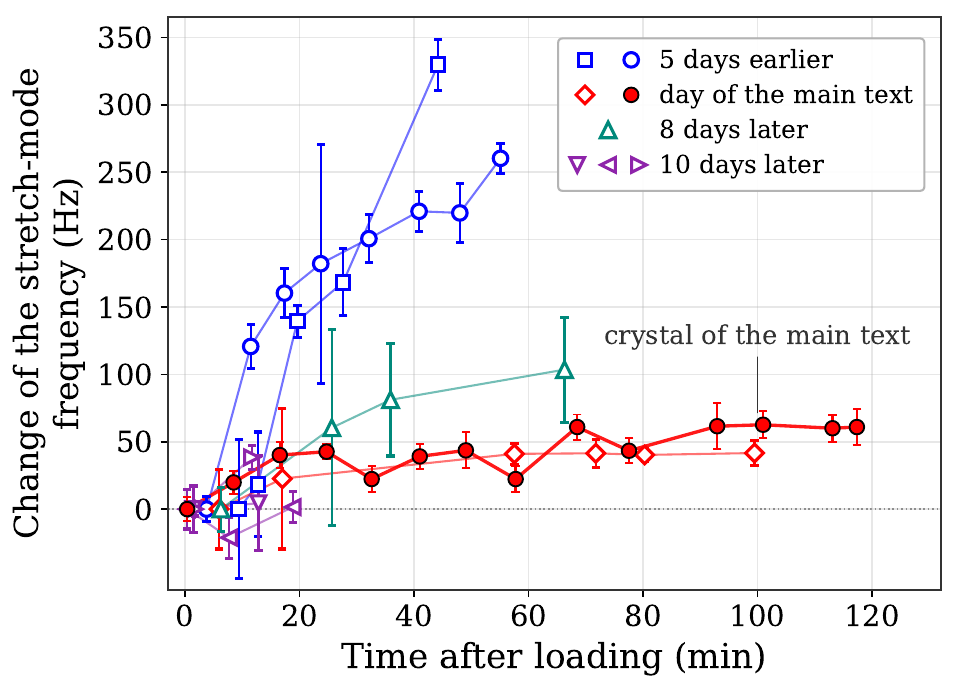}
\caption{Frequency of the stretch mode after a crystal is loaded, measured from the first sideband scan of each
crystal. Each point is one sideband scan. Markers: the different crystals of each day; the lines join the points of
one crystal.}
\label{fig:modedrift}
\end{figure}

\emph{Stability of the stretch mode.}---We measure the frequency of the stretch mode with sideband scans. A sideband
scan applies the gradient in two pulses of 1~ms each, at reduced amplitude, and records the probability of finding
one of the ions in its bright state. The frequency of the tone is stepped over $\pm3$~kHz around the mode from one
point of the scan to the next. Where the loops of the two pulses close, which is at every multiple of 1~kHz from the
mode, this probability keeps its starting value. Between these frequencies the spin is left entangled with the motion
and the probability moves towards $\tfrac12$. The envelope of this pattern falls off away from the mode, so that the
pattern is symmetric about the mode frequency only. We take its axis of symmetry as the mode frequency. The
uncertainty of this value has a median of 13~Hz over the 44 scans of Fig.~\ref{fig:modedrift}.

The mode frequency rises after loading, by an amount that differs from crystal to crystal
(Fig.~\ref{fig:modedrift}). Later than 40~min after loading it drifts by 23(8)~Hz per hour on the crystal of the
main text, and the scans of the settled crystals scatter about straight lines by no more than their uncertainties.
Beyond the drift, a random walk of the mode frequency, $\langle[\nu(t+\tau)-\nu(t)]^2\rangle=D\tau$, is therefore
limited to $D<0.23$~Hz$^2$/s, and flicker noise of spectral density $h_{-1}/f$ to $h_{-1}<27$~Hz$^2$ (95\% confidence for
each).

The DC electrodes are driven by DACs whose channels have a white noise floor of
60~nV/$\sqrt{\rm Hz}$, and at most 80~nV/$\sqrt{\rm Hz}$~\cite{Fastino}. In the potentials of the electrodes, computed
from the layout of the trap, one volt on an electrode moves the stretch mode by up to 10~kHz, and by 15~kHz in
quadrature over the twelve electrodes that set the well, so that independent noise of the channels moves the mode by
about 1~mHz/$\sqrt{\rm Hz}$. Noise of the DAC reference scales the voltages together and moves the mode by up to
about 2~mHz/$\sqrt{\rm Hz}$. Surface noise is a separate source.

\emph{Motional dephasing.}---Let the mode lie $\nu(t)$ above its calibrated frequency. In each
interval $dt$ the offset turns the displacement by $2\pi\nu\,dt$ in phase space and adds the energy
$h\nu|\alpha|^2\hat S_{z,m}^2$ to the displaced states. To first order this leaves
\begin{equation}
\delta\alpha=-2\pi i\!\int\!\nu(t)\,\alpha(t)\,dt,\qquad
\delta\Phi=2\pi\!\int\!\nu(t)\,|\alpha(t)|^2\,dt ,
\label{eq:mot:first}
\end{equation}
with the integrals taken over the whole gate. For loops of a constant force in one direction,
Eqs.~\eqref{eq:mot:err} and~\eqref{eq:mot:first} reproduce the rows of Ref.~\cite{Sutherland2022} for a static shift
and for white noise of the mode frequency exactly; the same holds for heating, below.

\begin{figure}[t]
\includegraphics[width=\columnwidth]{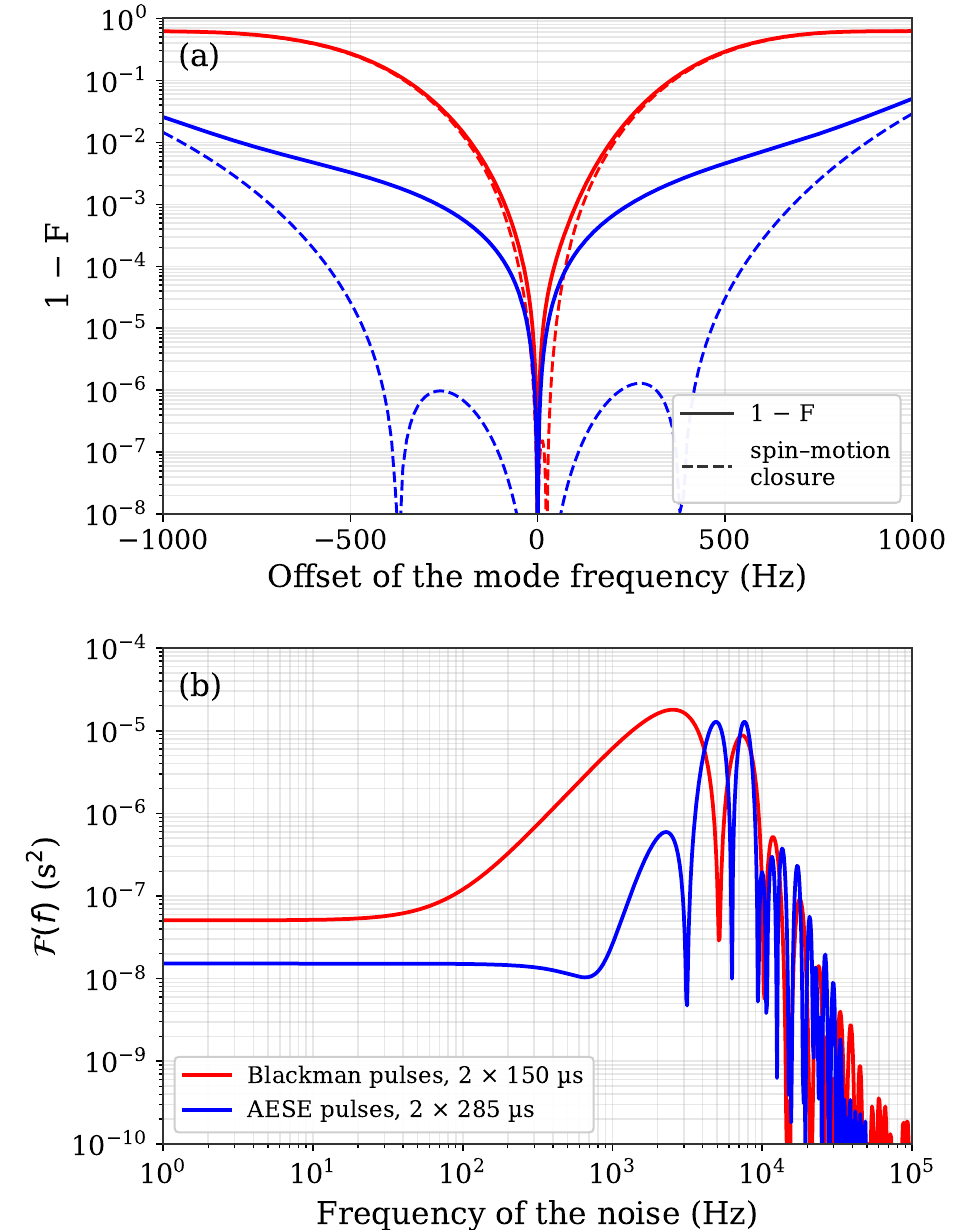}
\caption{Response of the two sequences to the frequency of the stretch mode, at $\bar n=50$. (a)~$1-F$ against
an offset of the mode frequency from its calibrated value that is constant during the gate. (b)~The error per unit
one-sided spectral density of the mode frequency, $\mathcal F(f)$, against the frequency of the noise, so that
$1-F=\int S(f)\mathcal F(f)\,df$.}
\label{fig:modenoise}
\end{figure}

For an offset $\nu$ that is constant during the gate, the first gradient pulse of the Blackman pair leaves a
displacement $\alpha_1$, which vanishes when the tone lies at the detuning at which a single pulse closes its loop,
7.226~kHz above the mode (Sec.~\ref{app:params:grad}). The second pulse leaves $-\alpha_1e^{i\theta}$, so that the pair
leaves
\begin{equation}
\alpha=\alpha_1\big(1-e^{i\theta}\big),\qquad \theta=2\pi\nu T_{\rm d},
\label{eq:mot:pair}
\end{equation}
where $T_{\rm d}=194~\mu$s is the time between the starts of the two pulses: of the displacement that one pulse
leaves, only the fraction $2\pi|\nu|T_{\rm d}$ remains. Each AESE pulse almost fully returns to the origin over a wide
range of offsets (Sec.~\ref{app:aese}). In both pairs the offset changes the geometric phase, by $2\pi\!\int\!|\alpha|^2dt$
per hertz: 0.11~mrad/Hz for the Blackman pair and 0.062~mrad/Hz for the AESE pair. Figure~\ref{fig:modenoise}(a) shows
$1-F$ against the offset, which is the odd population of the detuning scans of Fig.~3 of the main text
(Sec.~\ref{app:aese}).

For noise of the mode frequency with one-sided spectral density $S(f)$, Eq.~\eqref{eq:mot:first} gives
$1-F=\int S(f)\mathcal F(f)\,df$, where $\mathcal F(f)$ follows from the Fourier transforms of $\alpha(t)$ and
$|\alpha(t)|^2$ [Fig.~\ref{fig:modenoise}(b)]. Noise slower than about 100~Hz acts as an offset that is constant
during a gate, and the AESE pair
is 3.3 times less sensitive to it. Noise between about 0.3 and 5~kHz changes the mode frequency between the two
pulses. The Blackman pair then loses the cancellation of Eq.~\eqref{eq:mot:pair}, while the AESE pulses, which each
return to the origin, do not: at 1~kHz the AESE pair is 230 times less sensitive.

The noise of the DACs leads to an infidelity of less than $10^{-6}$ for either pair. For the surface
noise, our measured heating rate (below) bounds the differential electric field noise at the two ions at the mode
frequency; this is noise of the field gradient, below $9.4\times10^{-4}$~(V/m$^2$)$^2$/Hz. The gradient shifts the mode
frequency by 0.043~Hz per V/m$^2$, and we extrapolate the resulting $S(f)$ to lower frequencies as $1/f$. This leads to
an infidelity of at most $2\times10^{-5}$ for the Blackman pair and $6\times10^{-6}$ for the AESE pair.
These two make up the motional dephasing of Table~\ref{tab:motion}.

\emph{Quasi-static detuning error.}---The rest is the offset between the mode and its calibrated
frequency in each shot. It is constant during a gate, so it acts as an error of the detuning
[Fig.~\ref{fig:modenoise}(a)], and it grows with the time since the last calibration. Because the
mode moves fastest after loading, we took a sideband scan just before each of the two parity scans of the main text
and set the calibrated frequency with it, to about 9~Hz. A dedicated loading zone away from the gate zone would
largely remove this effect of loading. With the drift of the mode during each scan and a random walk or flicker
noise at its bound added, $1-F$ is below $1\times10^{-5}$ for each of the two scans. The same conditions give
$3\times10^{-6}$ for the AESE pair. Table~\ref{tab:motion} takes a typical calibration interval instead: a settled
crystal recalibrated every 20~min, with the drift of the crystal of the main text (23~Hz per hour), a random walk at
its bound and a calibration to 9~Hz. Averaged over the 20~min, $1-F$ is $1\times10^{-5}$ for the Blackman pair and
$4\times10^{-6}$ for the AESE pair.

\begin{table}[t]
\caption{Motional errors of the two sequences of the main text at $\bar n=50$ (this section). The
quasi-static detuning error is that of a settled crystal recalibrated every 20~min. The totals take the heating at
its bound.}
\label{tab:motion}
\begin{ruledtabular}
\begin{tabular}{lcc}
 & Blackman pair & AESE pair \\
\hline
Quasi-static detuning error & $1\times10^{-5}$ & $4\times10^{-6}$ \\
Motional dephasing & $2\times10^{-5}$ & $6\times10^{-6}$ \\
Anharmonicity & $7\times10^{-5}$ & $2\times10^{-5}$ \\
Field inhomogeneity & $2\times10^{-12}$ & $2\times10^{-12}$ \\
Heating ($\dot n<2$~s$^{-1}$) & $<1.4\times10^{-4}$ & $<8\times10^{-5}$ \\
\hline
Total & $2.4\times10^{-4}$ & $1.1\times10^{-4}$ \\
\end{tabular}
\end{ruledtabular}
\end{table}

\emph{Anharmonicity.}---The Coulomb interaction between the ions makes the stretch mode anharmonic. Let $d$ be
the distance between the ions and $u=u_0(a+a^\dagger)$ its change, with $u_0=(\hbar/M\omega_s)^{1/2}=10.9$~nm. The
expansion of the Coulomb energy $e^2/[4\pi\varepsilon_0(d+u)]$ in $\kappa=u_0/d=1.38\times10^{-3}$ adds, in units of
$\hbar\omega_s$, the terms $-\kappa\xi^3/6+\kappa^2\xi^4/6$, with $\xi=a+a^\dagger$. To second order in $\kappa$ the
energy levels of the mode become $\hbar\omega_sn+\hbar K(n^2+n)$ with $K=\omega_s\kappa^2/6$, so that the transition
frequency rises by $2K/2\pi=0.315$~Hz per phonon. In the same way the two out-of-phase radial modes shift it by
$-0.36$ and $-0.34$~Hz per phonon. The anharmonicity of the trap potential itself, computed from the potentials of the
electrodes, changes $K$ by less than $0.3\%$.

At $\bar n=50$ the mode frequency therefore differs from shot to shot by 16~Hz~rms, and the out-of-phase
radial modes, at the temperature of the stretch mode, add 9~Hz~rms. The self-Kerr term also makes the geometric phase depend on $n$:
the displaced states gain the energy $4\hbar Kn|\alpha|^2\hat S_{z,m}^2$, twice what the shift of the frequency alone would give. A
simulation of both sequences in the number basis, with the anharmonic mode, the calibrated frequency set to the mean of
the thermal state (as a sideband scan measures it) and the geometric phase calibrated, gives $1-F=6.3\times10^{-5}$ for the
Blackman pair and $1.6\times10^{-5}$ for the AESE pair. The out-of-phase radial modes add $5\times10^{-6}$ and $1\times10^{-6}$.

\emph{Field inhomogeneity.}---The force follows the gradient of the field at the ions. If the gradient changes
over the motion of the ions, the force depends on $n$~\cite{Sutherland2022}. For the field of the current-carrying
wires, computed with the Biot--Savart law from the layout of the chip, this gives $1-F=2\times10^{-12}$ for either
pair.

\emph{Heating.}---Heating adds random displacements to the motion while the states are displaced, and so
dephases them. For a gate of $N$ loops of a constant force in one direction, Eq.~(65) of Ref.~\cite{Sutherland2022}
gives, in our notation,
\begin{equation}
1-F=\frac{\pi\dot n}{4\Omega_gN^{1/2}}\,\lambda^2_{\hat S_{z,m}} .
\label{eq:mot:heat:su}
\end{equation}
Here $\dot n$ is the heating rate in quanta per second and
$\lambda^2_{\hat S_{z,m}}=\langle\hat S_{z,m}^2\rangle-\langle\hat S_{z,m}\rangle^2$ the variance of $\hat S_{z,m}$ in the
initial state of the qubits. Ref.~\cite{Sutherland2022} writes the gate Hamiltonian as Eq.~\eqref{eq:par:H}, with
$\Omega_{2g}$ for $\Omega_g/2$, and its loops close at $|\delta|=2\Omega_gN^{1/2}$.

The eigenstate of $\hat S_{z,m}$ with eigenvalue $s$ is displaced by $s\alpha(t)$. A random displacement $d\zeta$
of the mode at the time $t$ shifts the rest of the loop and adds the phase $2s\,\mathrm{Im}[\alpha^*(t)\,d\zeta]$ to this
state. Heating gives $\langle|d\zeta|^2\rangle=\dot n\,dt$, so the qubits acquire a random phase $\Theta\hat S_{z,m}$ with
$\langle\Theta^2\rangle=2\dot n\int|\alpha|^2dt$, which lowers the fidelity by
$\langle\Theta^2\rangle\lambda^2_{\hat S_{z,m}}$. For the loops of Ref.~\cite{Sutherland2022},
$\alpha=\Omega_g(1-e^{i\delta t})/2\delta$, and over the $N$ loops, which last $2\pi N/|\delta|$,
$\int|\alpha|^2dt=\pi/(8\Omega_gN^{1/2})$; this gives Eq.~\eqref{eq:mot:heat:su}. The argument holds for any
trajectory. In our gate the qubits start in an equal superposition of the four states, so
$\lambda^2_{\hat S_{z,m}}=2$ and
\begin{equation}
1-F=4\dot n\!\int\!|\alpha|^2dt ,
\label{eq:mot:heat}
\end{equation}
with the integral over the whole gate. It is 17.2~$\mu$s for the Blackman pair and 9.8~$\mu$s for the
AESE pair. We measured $\dot n<2$ quanta per second for the stretch mode, which gives $1-F<1.4\times10^{-4}$ and
$<7.9\times10^{-5}$.\space\space

\section{Magnetic dephasing}
\label{app:dephasing}

Magnetic-field noise shifts the qubit frequencies of the two ions together, since they are only a few micrometres
apart. We take its size from a single-ion spin echo and derive its effect on the parity contrast of the gate sequence
(Fig.~1 of the main text), in which it acts before the remapping pulse.

\begin{figure}[t]
\includegraphics[width=\columnwidth]{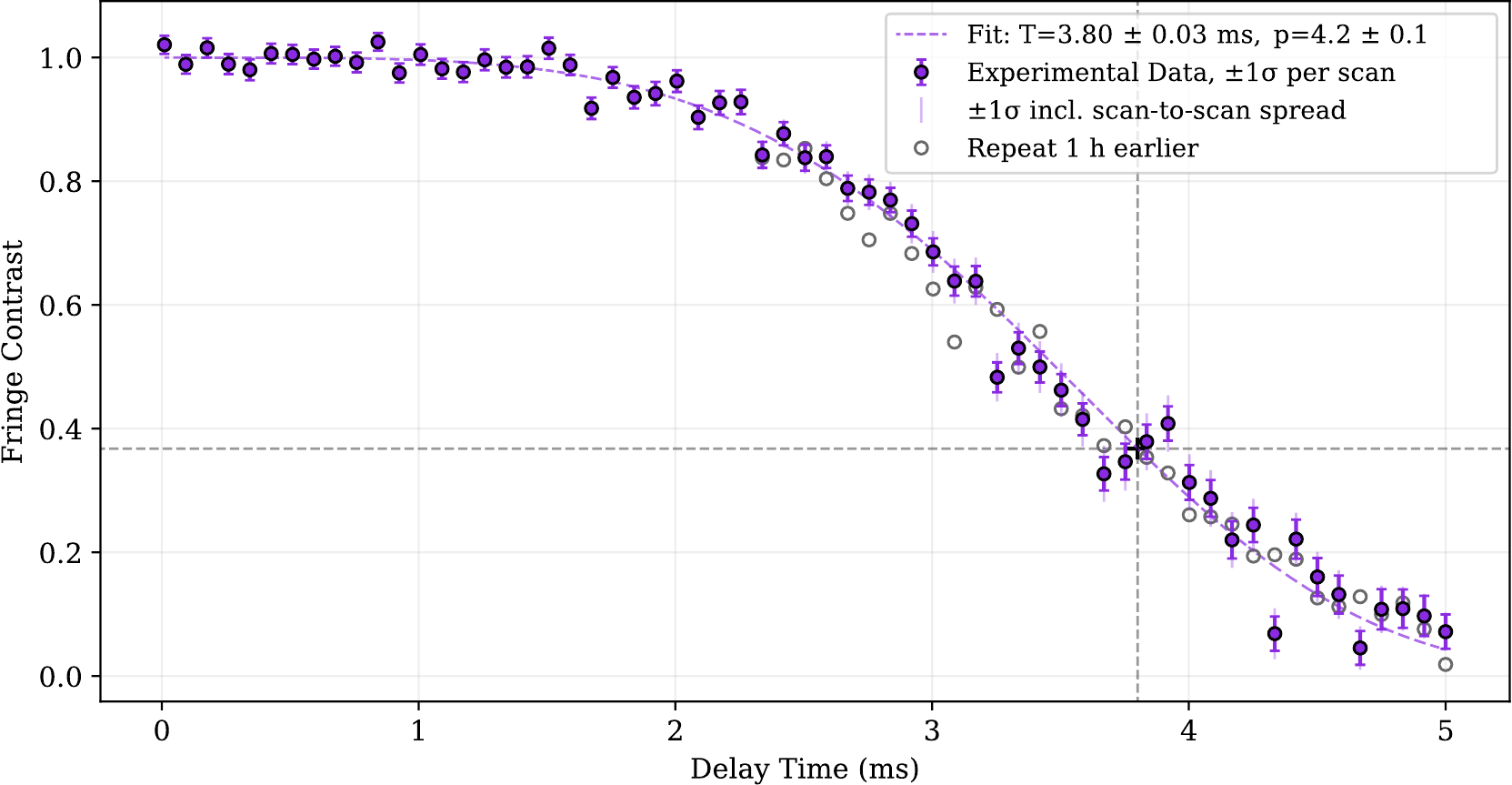}
\caption{Spin-echo contrast of a single ion on the gate qubit against the total delay $t$, each point from a scan
of the phase of the last pulse. The sequence is that of Fig.~1 of the main text without the gradient pulses and the
remapping pulse. The outer error bars include a scan-to-scan spread of 2.4\% in $T$. Dashed line:
Eq.~\eqref{eq:deph:echo}; the grey lines mark its $1/e$ point. The contrast is normalised to its mean at delays below
1~ms.}
\label{fig:decay}
\end{figure}

\emph{One ion.}---The contrast of a spin echo of a single ion on the gate qubit (Fig.~\ref{fig:decay}) decays with
the total delay $t$, with a decay time $T$ and an exponent $p$, as
\begin{equation}
C_1(t)=\exp\!\big[-(t/T)^p\big],\qquad T=3.80(3)~\mathrm{ms},\quad p=4.2(1).
\label{eq:deph:echo}
\end{equation}
An exponent near four is expected for noise that is slow compared with the sequence: the echo cancels a constant
field, and a field that changes linearly in time leaves a phase that grows as $t^2$. For a Gaussian phase $\beta$ of
variance $\sigma_\beta^2$ the contrast is $\langle\cos\beta\rangle=e^{-\sigma_\beta^2/2}$, so that
\begin{equation}
\sigma_\beta^2(t)=2\,(t/T)^p .
\label{eq:deph:var}
\end{equation}

\emph{Two ions.}---Both ions acquire the same phase, an error $Z(\beta)\otimes Z(\beta)$ with
$Z(\beta)=e^{-i\beta\sigma_z/2}$. Because it commutes with the $\sigma_z\sigma_z$ interaction, the phase acquired
between the first and the remapping pulse acts as a single error just before the remapping pulse, with the sign of
its part before the echo pulse reversed. This is the phase of a single-ion echo of the same length $t$, with the
variance of Eq.~\eqref{eq:deph:var}. Just before the $\pi/2$ remapping pulse the ideal state is
\begin{equation}
|\psi\rangle=\tfrac12\big(|00\rangle-|01\rangle-|10\rangle-|11\rangle\big),
\label{eq:deph:psi}
\end{equation}
where $|0\rangle$ and $|1\rangle$ are the two states of the gate qubit. The error leaves the odd-parity components
$|01\rangle$ and $|10\rangle$ unchanged, a decoherence-free subspace for common phase noise, and changes the relative
phase of $|00\rangle$ and $|11\rangle$ by $2\beta$. After the remapping pulse and the parity pulse of phase $\phi$,
the parity is
\begin{equation}
\Pi(\phi)=\tfrac12\sin^2\!\beta\,(1-\cos2\phi)-\cos\beta\,\sin2\phi .
\label{eq:deph:parity}
\end{equation}
Its fringe at $2\phi$ has the offset $c=\tfrac12\sin^2\!\beta$ and the contrast
\begin{equation}
C(\beta)=\big(\cos^2\!\beta+\tfrac14\sin^4\!\beta\big)^{1/2}=1-\tfrac12\sin^2\!\beta ,
\label{eq:deph:Cbeta}
\end{equation}
and its phase moves only at second order in $\beta$: the error adds even-parity outcomes at the minima of the fringe
and leaves its maxima unchanged.

\emph{Average over shots.}---A measured fringe is the average over many shots, so its contrast is that of the averaged
Fourier components. With $\langle\cos k\beta\rangle=e^{-k^2\sigma_\beta^2/2}$ these are
$-\tfrac12\langle\sin^2\beta\rangle=-\tfrac14\big(1-e^{-2\sigma_\beta^2}\big)$ and
$-\langle\cos\beta\rangle=-e^{-\sigma_\beta^2/2}$, and
\begin{equation}
C(\sigma_\beta)=\Big[e^{-\sigma_\beta^2}+\tfrac1{16}\big(1-e^{-2\sigma_\beta^2}\big)^2\Big]^{1/2}.
\label{eq:deph:C}
\end{equation}
For small errors $C\simeq1-\sigma_\beta^2/2=1-(t/T)^p$. To first order the two-ion parity contrast thus loses as much
as the single-ion echo contrast, $C_1\simeq1-(t/T)^p$, and four times less than a Bell state
$|00\rangle+e^{i\chi}|11\rangle$ exposed to the same noise, whose parity contrast is
$e^{-2\sigma_\beta^2}\simeq1-2\sigma_\beta^2$. For large errors $C\to\tfrac14$, the fringe of the coherence between
$|01\rangle$ and $|10\rangle$, which the noise does not touch.

\emph{The gate.}---From its first microwave pulse to the remapping pulse the gate sequence lasts $t\approx380$~$\mu$s: the two gradient
pulses of 150~$\mu$s, the microwave pulses of the echo and the delays around the gradient pulses. Magnetic dephasing then reduces the
parity contrast by $(t/T)^p=7(2)\times10^{-5}$, where the uncertainty is that of the exponent; this is the value
of the error budget. The AESE sequence of Sec.~\ref{app:aese} lasts $t\approx0.64$~ms, for which the loss is
$6(1)\times10^{-4}$.

\section{Differential amplitude jitter of the gradient pulses}
\label{app:jitter}

The transverse field of the oscillating-gradient drive shifts the qubit frequency through the AC Zeeman effect, and
each gradient pulse imprints a phase $\Phi_{\rm Z}$${}\approx4900$~rad on each ion (Sec.~\ref{app:params:acz}). The spin echo
cancels this phase only as far as the two pulses have the same amplitude. Since $\Phi_{\rm Z}$ grows about as the square of the amplitude, a difference
$\Delta a$ between the fractional amplitudes of the two pulses leaves each ion with the phase
$\beta=2\Phi_{\rm Z}\,\Delta a$, nearly the same for both ions.

\emph{Noise.}---We model the fractional amplitude noise of the drive chain by the one-sided spectral density
\begin{equation}
S_a(f)=h_0+h_{-1}/f
\label{eq:jit:psd}
\end{equation}
at the frequency $f$. The white term is thermal noise: a single-sideband AM noise of
$\mathcal{L}_{\rm AM}=-165$~dBc/Hz gives $h_0=2\mathcal{L}_{\rm AM}=6.3\times10^{-17}$~Hz$^{-1}$. The flicker term we
estimate from a typical AM noise of $\mathcal{L}_{\rm AM}=-140$~dBc/Hz at $f=1$~kHz: $h_{-1}=2f\mathcal{L}_{\rm AM}
=2\times10^{-11}$.

\emph{Filter.}---The echo is sensitive to $\Delta a=\bar a_1-\bar a_2$, where $\bar a_i$ is the fractional amplitude
averaged over pulse $i$ with the weight of the AC Zeeman shift, the squared pulse envelope. Its variance is
\begin{equation}
\langle\Delta a^2\rangle=\int_0^\infty S_a(f)\,4\,|W(f)|^2\sin^2(\pi fT_{\rm d})\,df,
\label{eq:jit:filter}
\end{equation}
where $W(f)$ is the Fourier transform of that weight, normalised to unit area, and $T_{\rm d}$ the time between the
starts of the two pulses. The filter vanishes as $f^2$ at low frequencies, so that the flicker term converges, and
the smooth edges make it fall as $f^{-6}$ at high frequencies. For two pulses of 150~$\mu$s with 10-$\mu$s Blackman
edges, 44~$\mu$s apart ($T_{\rm d}=194~\mu$s), Eq.~\eqref{eq:jit:filter} gives
$\langle\Delta a^2\rangle^{1/2}\approx8.5$~ppm, almost all of it from the flicker term; the white term alone gives
0.7~ppm. With the specialised electronic circuit of the drive chain the jitter is
$\langle\Delta a^2\rangle^{1/2}\approx1.1$~ppm.

\emph{Contrast.}---With $\Phi_{\rm Z}\approx4900$~rad the phase left on each ion has the standard deviation
$\sigma_\beta=2\Phi_{\rm Z}\langle\Delta a^2\rangle^{1/2}\approx11$~mrad. This is a common phase error of the kind treated in
Sec.~\ref{app:dephasing}, and it reduces the parity contrast by $\sigma_\beta^2/2\approx6\times10^{-5}$; for so
small a phase this holds whatever the distribution of $\beta$.

\section{Third harmonic of the tone and the radial modes}
\label{app:radial}

The third harmonic of the tone, at 1.521~MHz, lies 57~kHz above the upper radial mode, and a spin-dependent force at
that frequency would entangle the spins with the radial motion without closing a loop. Such a force has two sources.
The first is a third harmonic of the drive current. It is at least
60~dB below the fundamental, which limits its gradient at the ions to 0.04~T/m and its error to $10^{-7}$. The
second is the qubit itself: its frequency follows the magnitude of the field, which is not linear in the drive when the
transverse field is comparable to the static field (Sec.~\ref{app:params:acz}), so that a purely sinusoidal current
gives a force at the harmonics of the tone.

To estimate it we continue the field measured on the trap axis (Sec.~\ref{app:params:acz}) away from the axis with
Maxwell's equations: the wires cross the axis in the plane of the chip, so their field does not depend on the radial
coordinate in that plane, and its vertical component then changes with height at the rate of the axial gradient. The
vertical gradient of the field magnitude at the two ions has a time average of 10--12~T/m and components of
0.7--2.7~T/m at the tone's frequency, 10--11~T/m at twice and 0.2--0.3~T/m at three times that frequency. The time
average acts as a static gradient (AC Zeeman shift gradient): it shifts the ions by a few picometres, depending on their spin, and changes the
geometric phase by $10^{-4}$~rad.

We integrate the gate sequence in the time domain with this gradient and with the four radial modes of the
crystal (the in-phase modes of Table~\ref{tab:params} and the out-of-phase modes at 1.354 and 1.435~MHz) at the
temperature of the stretch mode, which is an occupation of 17 for them. The error is $3\times10^{-6}$ if the upper radial axis is vertical, the worst orientation, and
$1.5\times10^{-6}$ at $45^\circ$. It oscillates with the distance between the third harmonic and the mode, through the
spectrum of the pulses, and reaches $1\times10^{-5}$ if that distance changes by a few kilohertz; the uncertainty
of the transverse gradient of the field allows $2\times10^{-5}$. We enter $2\times10^{-5}$ in the error budget.

\section{Leakage out of the qubit}
\label{app:leakage}

The qubit states $|1,\pm1\rangle$ are coupled to $|1,0\rangle$ by magnetic fields that oscillate near the Larmor
frequency. The strongest such field at the ions is that of the trap drive, 0.25~G at 15.70~MHz, perpendicular to the
trap axis; we measured it from the splitting of the microwave line when the Zeeman splitting was tuned to the trap
frequency. During a gradient pulse the Larmor frequency follows the magnitude of the total field and sweeps between
23.3 and 30.5~MHz in each period of the tone. One quantum of the trap drive lies
7.6~MHz below this range and two quanta, at 31.40~MHz, lie 0.9~MHz above it, so that leakage is a two-photon process
in the wing of the comb of lines that the modulation makes of each transition.

We integrate the evolution of the $F=1$ manifold through one gradient pulse numerically, with the drive field of
Sec.~\ref{app:params:acz} at the positions of the two ions and the trap field added. The two-photon process needs one
quantum polarised across the instantaneous quantisation axis and one along it, and the gradient drive tilts the axis
only within the plane of the static field and its own transverse field. By the symmetry of the electrodes the trap
field lies in the plane of the chip and the transverse field of the gradient drive points out of it, so that the two
are perpendicular and the process is suppressed. For this geometry the two pulses of the gate leave
$4\times10^{-8}$ in $|1,0\rangle$ at the static field used, most of it on the ion whose Larmor frequency comes
closest to twice the trap frequency. We enter $4\times10^{-8}$ in the error budget.

\section{The AESE pulses: robustness to the detuning and contrast}
\label{app:aese}

\begin{figure}[t]
\includegraphics[width=\columnwidth]{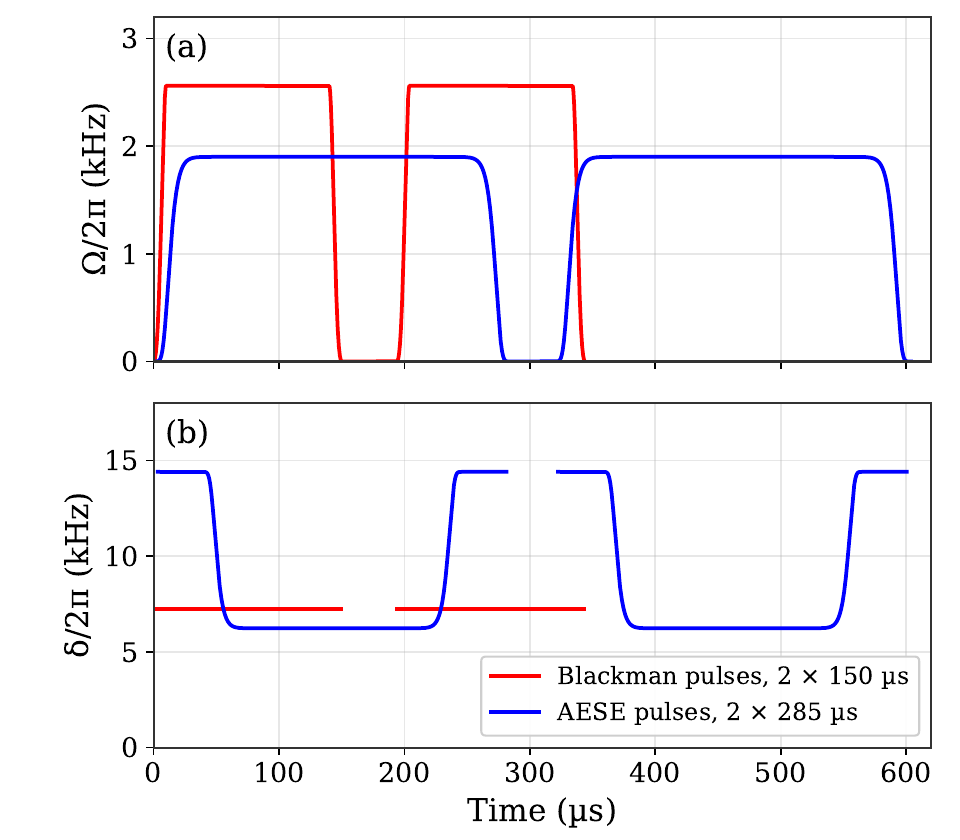}
\caption{The two pulse pairs of the main text to scale, the detuning $\delta$ shown while each pulse is on. The microwave
echo pulses lie in the gaps between the two pulses of each pair.}
\label{fig:pulses}
\end{figure}

\emph{The pulses.}---Each of the two AESE pulses has a super-Gaussian amplitude and a super-Gaussian dip of the
detuning,
\begin{align}
\Omega_g(t)&=\Omega_{\max}\exp\!\big[-\ln2\,|2t/w_\Omega|^{p_\Omega}\big],\nonumber\\
\delta(t)&=\delta_{\max}-(\delta_{\max}-\delta_{\min})\exp\!\big[-\ln2\,|2t/w_\delta|^{p_\delta}\big],
\label{eq:aese:pulse}
\end{align}
with $t$ counted from the centre of the pulse, full widths at half maximum $w_\Omega$ and $w_\delta$ and exponents
$p_\Omega$ and $p_\delta$. The force is switched on while the tone is far from the mode, at
$\delta_{\max}$; the tone is then brought to $\delta_{\min}$ and back before the force is switched off, so that the
motion follows the force and returns to its starting point for a range of detunings. Each pulse closes its own loop
and contributes a geometric phase of $|\Phi|=\pi/16$. The pulses were about 285~$\mu$s long, with $w_\Omega=259~\mu$s,
$p_\Omega=32.0$, $w_\delta=185~\mu$s, $p_\delta=25.8$, $\delta_{\max}=-2\pi\times14.4$~kHz,
$\delta_{\min}=-2\pi\times6.2$~kHz and a peak gate Rabi frequency of $\Omega_{\max}=2\pi\times1.9$~kHz, for which this
pulse shape gives the measured geometric phase. The amplitude rises from
10\% to 90\% in 12~$\mu$s and stays within 1\% of its peak for 226~$\mu$s; the detuning changes from 10\% to 90\% of
its dip in 11~$\mu$s and stays within 1\% of $\delta_{\min}$ for 157~$\mu$s. The two pulses were 34~$\mu$s apart.
Figure~\ref{fig:pulses} shows them to scale, together with the Blackman pulses of the gate.

These pulses were run at a lower drive than the gate of Table~\ref{tab:params}, with the stretch mode at
488.6~kHz. The geometric phase is measured by scanning the phase of the
$\pi/4$ pulse of Sec.~\ref{app:params:nbar}. At the same drive, Blackman-edged pulses reach the full geometric phase with
180~$\mu$s each (the same measurement gave $1.02(4)\times\pi/8$), a gate Rabi frequency of 2.1~kHz in the model of
Sec.~\ref{app:params:grad}. At equal drive the AESE pulses are thus 1.6 times longer than the Blackman pulses.

\emph{Model.}---Let the mode lie $\nu$ above its calibrated frequency, as in Sec.~\ref{app:motion}. The
detuning error of Fig.~3 of the main text is then $-\nu$. One pulse leaves the displacement $\alpha_1$ of
Eq.~\eqref{eq:par:alpha} with $\delta\to\delta+2\pi\nu$, and the pair leaves $\alpha=\alpha_1(1-e^{i\theta})$ with
$\theta=2\pi\nu T_{\rm d}$ (Eq.~\eqref{eq:mot:pair}) and the phase $\Phi=2\Phi_1-|\alpha_1|^2\sin\theta$, with
$\Phi_1$ the phase of one pulse and $T_{\rm d}$ the time between the starts of the pulses (319~$\mu$s for the AESE
pulses). The force is fixed by
$|\Phi|=\pi/8$ at $\nu=0$. In a detuning scan
(Sec.~\ref{app:params:nbar}) the population of the odd outcomes is
\begin{equation}
P_{\rm odd}=\tfrac12\Big[1-e^{-G}\cos\!\big(4|\Phi|-\tfrac{\pi}{2}\big)+\tfrac14\big(1-e^{-4G}\big)\Big],
\label{eq:aese:podd}
\end{equation}
with $G=4(\bar n+\tfrac12)|\alpha|^2$, which rises from zero to $5/8$ as the motion becomes entangled with the spins,
and the fidelity of the Bell state is $F=[6+2e^{-4G}+8e^{-G}\cos(4|\Phi|-\pi/2)]/16$, the expression of which
Eq.~\eqref{eq:mot:err} is the expansion. The two are related by $P_{\rm odd}=1-F$: the odd population of such a scan
is the infidelity of the gate, as far as a displacement left on the mode and an error of the geometric phase cause it.
Fig.~3 of the main text shows $1-P_{\rm odd}$.

\emph{Results.}---The lines of Fig.~3 of the main text are this model for the occupation of
Sec.~\ref{app:params:nbar}, $\bar n=50$, with the frequency of the mode as the only fitted parameter of each scan.
$P_{\rm odd}$ stays below 1\% (the fidelity above 99\%) for detuning errors between $-680$ and $+795$~Hz
for the AESE pulses, and between $-185$ and $+180$~Hz for Blackman-edged pulses of 152.5~$\mu$s: a range four times
as wide. The model describes the AESE scan
with $\chi^2=10.4$ for 8 degrees of freedom (nine detunings, one fitted parameter); with the occupation fitted as
well, this scan gives $\bar n=54(4)$. The scan of the Blackman pulses was taken in an earlier session. Its points lie
below the model between 250 and 500~Hz from the centre, and the scan is better described by a higher occupation,
$\bar n=83(4)$. This crystal was hotter than the $\bar n\approx50$ of the other data sets
(Sec.~\ref{app:params:nbar}), because excess micromotion was left uncompensated in that session: it
broadens the cooling transition as the ions see it, and so raises the temperature that Doppler cooling reaches.
The inset of that figure shows the paths $\alpha(t)$ of the two pulses in phase space for a detuning
error of 300~Hz: the first Blackman pulse ends at $|\alpha_1|=0.047$, which the second pulse, drawn as a
faint line, reduces to $|\alpha|=0.017$ ($G=0.06$), and each AESE pulse ends within $10^{-3}$ of the
origin.

Within its flat region the AESE pair leaves almost no displacement, $G<10^{-4}$ up to 500~Hz, against
$8\times10^{-4}$ at 100~Hz for the Blackman pulses. Its error is that of the
geometric phase, which changes by 1.6\% per 100~Hz, against 2.8\% for the Blackman pulses. A scan of the detuning with the
$\pi/2$ remapping pulse, which measures the geometric phase through the difference of the two even populations, gives
1.6\% per 100~Hz. The gate error $1-F$ therefore stays below $10^{-3}$ between $-245$ and $+265$~Hz, against
$-95$ to $+90$~Hz for the Blackman pulses, for which the displacement dominates. For Blackman pulses of 180~$\mu$s,
those of the drive at which the AESE pulses ran, the model gives $P_{\rm odd}<1\%$ between $-150$ and $+155$~Hz and
a gate error below $10^{-3}$ between $-75$ and $+80$~Hz: at equal drive the two ranges of the AESE pulses are 4.8
and 3.3 times wider. Section~\ref{app:motion} compares the two pairs for noise of the mode frequency and for
the other motional errors.

\emph{Contrast.}---The parity scan of the AESE sequence in Fig.~2 of the main text has 1900 shots at each
of its nine phases. Its readout is calibrated on 13 single-ion SPAM runs of the preceding night, with
$1.25\times10^6$ shots per state in all, taken with the detection light switched on 0.4~ms after the camera's trigger
(Sec.~\ref{app:spam:model}). These runs give the parameters of the camera and the failure probabilities at the start
of the window, $\epsilon^{\rm B}=0.09(2)\%$ and $\epsilon^{\rm D}=0.007(4)\%$, which are used for both ions. An ion
is pumped dark at a rate proportional to the rate at which it scatters light, so that the pumping rate per detected
photo-electron is the same for every ion read with this camera; between the ions for which we measured it, it
differed by 7\%. Each ion's pumping rate is therefore set to its own detected light times the value of the single-ion
runs, $\alpha_{\rm bd}/\lambda=0.0210(5)$.

The contrast is obtained from a fit of the fringe to all shots. It gives
$C=1.0043\pm0.0058\,(\mathrm{stat})\pm0.0033\,(\mathrm{readout})$ with $c=+0.003\pm0.005$, which the main text gives
with three decimals. The readout error is that of the pumping rates, $\pm0.0030$, of the failure probabilities,
$\pm0.0013$, and of the camera's parameters, $\pm0.0003$. By the construction of Sec.~\ref{app:spam:gate}, with
$\sigma_C=0.0071$, the interval runs from 0.9965 to 1, $F=100.0^{+0.0}_{-0.3}\%$.